\documentclass[prc,superscriptaddress,nofootinbib,noeprint,dvipsnames]{revtex4-2}

\usepackage[english]{babel}
\usepackage[hidelinks]{hyperref}
\usepackage{dcolumn}
\makeatletter\AtBeginDocument{\let\@elt\relax}\makeatother

\usepackage{booktabs}
\usepackage{longtable}
\usepackage{tabularx}
\usepackage{graphicx}
\usepackage{amsmath}
\usepackage{float}
\usepackage[english,capitalise]{cleveref}
\usepackage{mathtools}
\usepackage{bbm}

\usepackage[utf8]{inputenc}

\usepackage{siunitx}

\usepackage{physics}
\usepackage{csquotes}
\usepackage{multirow}
\usepackage{ifthen}
\usepackage{tikz}
\usetikzlibrary{calc}
\usetikzlibrary{decorations.pathmorphing}
\usepackage{varwidth}

\usepackage{pifont}
\newcommand{\checkm}{\ding{51}}
\newcommand{\crossm}{\ding{55}}

\makeindex

\usepackage{our_newcommands0722}

\newcommand{\orcid}[1]{\href{https://orcid.org/#1}{\includegraphics[scale=0.055]{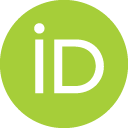}}}

\definecolor{lgreen1}{RGB}{140, 245, 140}
\definecolor{lblue1}{RGB}{90, 170, 255}
\definecolor{blue1}{RGB}{65,135,235}
\definecolor{orange1}{RGB}{235,155,70}
\definecolor{aqua1}{RGB}{130, 220, 230}

\newcommand{\dnc}{blue1} 
\newcommand{\dcc}{orange1} 
\newcommand{\dncc}{lgreen1} 
\newcommand{\dnnc}{lblue1} 
\newcommand{\dtbv}{} 
\newcommand{\danc}{lgreen1} 
\newcommand{\dacc}{lblue1} 

\newcommand{\daspar}{0.6} 

\newcommand{\lieleven}{\({}^{11}\mathrm{Li}\)}
\newcommand{\liten}{\({}^{10}\mathrm{Li}\)}
\newcommand{\linine}{\({}^{9}\mathrm{Li}\)}

\newcommand{\mofod}{\K{\Omega^{(\mathrm{fo})}}^\dagger} 

\newcommand{\pcz}{P_c^{(0)}}
\newcommand{\pco}{P_c^{(1)}}
\newcommand{\pnm}{P_n^{(-)}}
\newcommand{\pnp}{P_n^{(+)}}

\newcommand{\xicmi}[1]{ \xi_c^{\K{#1;J,M}} }
\newcommand{\xinmi}[1]{ \xi_n^{\K{#1;J,M}} }

\begin{document}

\title{\texorpdfstring{Final-state interactions and spin structure in $\boldsymbol{E1}$ breakup of
  \(\boldsymbol{^{11}}\)Li in Halo EFT}{Final-state interactions and spin structure in E1 breakup of
  Li-11 in Halo EFT}}


\author{Matthias G\"{o}bel \orcid{0000-0002-7232-0033}}
\email[E-mail: ]{goebel@theorie.ikp.physik.tu-darmstadt.de}
\affiliation{Technische Universit\"{a}t Darmstadt, Department of Physics, Institut f\"{u}r Kernphysik, 64289 Darmstadt, Germany}

\author{Bijaya Acharya \orcid{0000-0003-1192-093X}}
\affiliation{Physics Division, Oak Ridge National Laboratory, Oak Ridge, TN 37831, USA}
\affiliation{Institute of Nuclear and Particle Physics and Department of Physics and Astronomy, Ohio University, Athens, OH 45701, USA}

\author{Hans-Werner Hammer \orcid{0000-0002-2318-0644}}
\affiliation{Technische Universit\"{a}t Darmstadt, Department of Physics, Institut f\"{u}r Kernphysik, 64289 Darmstadt, Germany}
\affiliation{ExtreMe Matter Institute EMMI and
Helmholtz Research Academy for FAIR, GSI Helmholtzzentrum f\"{u}r Schwerionenforschung GmbH, 64291 Darmstadt, Germany}

\author{Daniel R. Phillips \orcid{0000-0003-1596-9087}}
\affiliation{Institute of Nuclear and Particle Physics and Department of Physics and Astronomy, Ohio University, Athens, OH 45701, USA}


\date{\today}

\begin{abstract}
  We calculate the $E1$ breakup of the \(2n\) halo nucleus \(^{11}\)Li in
  Halo Effective Field Theory (Halo EFT) at leading order. In Halo EFT, \(^{11}\)Li
  is treated as a three-body system of a \(^{9}\)Li core and two neutrons. 
  We present a detailed investigation of final-state interactions (FSI) in the
  neutron-neutron \((nn)\) and neutron-core \((nc)\) channels.
We employ M{\o}ller operators to formulate an expansion scheme that satisfies the non-energy-weighted cluster sum rule and successively includes higher-order terms in the multiple-scattering series for the FSI. Computing the $E1$ strength up to third order in this scheme, we observe apparent convergence and good agreement with experiment. 
The neutron-neutron FSI is by far the most important contribution and largely determines the maximum value of the $E1$ distribution. However, inclusion of \(nc\) FSI
does shift the peak position to slightly lower energies.
Moreover, we investigate the sensitivity of the $E1$ response to the spin structure of the neutron-${}^9$Li interaction. We contrast results for an interaction that is the same in the spin-1 and spin-2 channels with one that is only operative in the spin-2 channel, and find that good agreement with experimental data is only obtained if the interaction is present in both spin channels. The latter case is shown to be equivalent to a calculation in which the spin of $^9$Li
is neglected.
\end{abstract}
\maketitle


\section{Introduction}

Halo nuclei consist of a compact core and one or more loosely
bound valence nucleons. As a consequence, they are significantly
larger than neighbouring nuclei in their isotopic chain.
Neutron halos are the most universal halo systems as their halo structure
is not altered by the long-range Coulomb
interaction~\cite{Riisager:2012it,Hammer:2017tjm}.
They were discovered in the 1980's at radioactive beam
facilities by measuring their unusually large interaction
radius~\cite{Tanihata:2016zgp}. Jonson and Hansen
subsequently showed that this
large radius is connected to a small separation energy of the halo
neutrons~\cite{Hansen:1987mc}.

The corresponding separation of energy scales forms the basis for a controlled
description of halo nuclei in the framework of Halo
Effective Field Theory (Halo EFT) \cite{Bertulani:2002sz,Bedaque:2003wa,Hammer:2017tjm},
which systematizes cluster models of halo nuclei.
The breakdown scale $M_{core}$ is the 
lowest momentum scale not explicitly included in
the theory. This is set by the excitation energy of
the core, or by the size of the core, whichever yields the smaller momentum scale.
The EFT exploits that the momentum scale of the halo nucleons
set by their separation energy is much smaller, $M_{halo}\ll M_{core}$.
Typically $M_{halo}$ is of order tens of MeV for halo nuclei, while,
the breakdown momentum scale, $M_{core}$, varies between 50 and 150 MeV.
The EFT expansion is then in powers of $M_{halo}/M_{core}$,
and for a typical momentum of order $M_{halo}$ the EFT uncertainty
is of order $(M_{halo}/M_{core})^{n+1}$  for a calculation at order $n$.
Halo EFT describes the structural properties
of one- and two-neutron halo nuclei with nucleon-nucleon and nucleon-core interactions.
It has has also been applied to a number of electromagnetic and
weak observables, including capture reactions,
photodissociation processes, and weak
decays (see, e.g., Refs.~\cite{Hammer:2017tjm,Higa:2016igc,Premarathna:2019tup,Zhang:2019odg,Elkamhawy:2019nxq} for a review and some recent references.)

In this work, we focus on
Coulomb dissociation which is a powerful tool to study the structure
of halo nuclei. 
The electric dipole transition strength, which is enhanced at low excitation energies for halo nuclei, is probed in Coulomb dissociation experiments by accelerating them to high energies and scattering them peripherally off a high-Z target. 
This ``soft dipole mode'' has been under intense
investigation both in experiment and in theory since the discovery
of halo nuclei in the early 80's~\cite{Aumann_2013}.
Halo EFT was first applied to Coulomb dissociation of the
one-neutron halo $^{11}$Be \cite{Hammer:2011ye}. Further work
extended the description to $^{19}$C~\cite{Acharya:2013nia}.
In Ref. \cite{Hammer:2017tjm}, it  was shown that the $E1$ excitation
of one-neutron halo nuclei can be described by a dimensionless universal
function of the energy in units of the one-neutron separation energy.

The \(E1\) response of a two-neutron halo is also expected to be governed by a universal
function \cite{Baur:2006ve}, in close analogy to the one-neutron case discussed above.
In this work, we discuss the $E1$ response of $^{11}$Li. Lithium-11 was previously
considered in Halo EFT in Refs.~\cite{Canham:2008jd,Canham:2009xg,Hagen:2013jqa,Vanasse:2016hgn}.
Both $^{11}$Li and the $^{9}$Li core have the quantum number $J^P = 3/2^-$,
while the unbound $^{10}$Li appears to have a low-energy anti-bound state with quantum numbers $J^P = 2^-$ or $1^-$.

The two-neutron separation energy of $^{11}$Li is 0.369 MeV and the $^{10}$Li \(s\)-wave resonance\footnote{
  Due to the mentioned inconclusive state of the literature this resonance is in many cases also understood as a virtual state. 
  Moreover, there is some discussion whether \liten~can be a low-energy \(s\)-wave resonance.
  In principle the effective nuclear \linine-\(n\) interaction could form a barrier. Thereby \liten could be a true resonance.
} is
26(13) keV above the \linine-\(n\) threshold \cite{Wang:2021xhn}.
The momentum scale $M_{halo}$ can be estimated
from these energy scales using $M\sim\sqrt{mE}$ as $M_{halo}=18.6$ MeV. 
The first excitation energy of
the $^9$Li ground state is 2.69 MeV \cite{Tilley:2004zz} and its one-neutron separation energy
is 4.06 MeV \cite{Wang:2021xhn},
while the charge radius of ${}^9$Li is 2.25 fm~\cite{Nortershauser:2011zz},
implying a scale $M_{core}$ of 50--90 MeV.
This yields an expansion parameter of, at worst, $M_{halo}/M_{core}\sim 0.37$.

A leading-order Halo EFT calculation of ${}^{11}$Li should therefore be able to describe Coulomb dissociation data with reasonable accuracy. This was indeed found to be the case in
preliminary Halo EFT calculations of this process~\cite{Hagen:2014,Acharya:2015,Acharya:2015gpa} that showed good agreement with the $E1$ strength
extracted from Coulomb dissociation data in Ref.~\cite{Nakamura:2006zz} at
transition energies within the domain of validity of Halo EFT.
These studies, as well as earlier work  within
three-body models in Refs.~\cite{Esbensen:1992qbt,Esbensen:2007ir,Kikuchi:2013ula} found that, in contrast to the case of Coulomb dissociation
of an \(s\)-wave one-neutron halo~\cite{Acharya:2015}, final-state interactions (FSIs) play a significant
role in determining the neutron spectrum measured in Coulomb dissociation of ${}^{11}$Li.
This was also observed for other \(2n\) halos such as \({}^{22}\)C, see, e.g., Ref. \cite{Ershov:2012fy}.

In this work, our aim is threefold:
\begin{itemize}
\item Obtain a description of the experimental $E1$ breakup data from
  Ref.~\cite{Nakamura:2006zz} in Halo EFT with theoretical uncertainties.
\item Explore in detail the role of the $nn$ and \linine-\(n\) FSIs in this process, paying particular attention to the constraints from the non-energy-weighted
  cluster sum rule for $E1$ breakup.
\item Investigate the impact of the spin structure of the
  \linine-\(n\) interaction, which is non-trivial because the ${}^9$Li
  core has spin $3/2$, and derive the relationship of such a calculation to the frequent assumption of a spin-0 core.
\end{itemize}

As was done in Refs.~\cite{Canham:2008jd,Canham:2009xg,Hagen:2013jqa,Vanasse:2016hgn}, we include only \(s\)-wave interactions in our leading-order calculation.
We are aware that this is a different strategy to the one typically taken in three-body cluster models, many of which predict significant \(p\)-wave components in
the wave function~\cite{Zhukov:1993aw}.
There is also experimental evidence for a mixing of different-parity components \cite{Simon:1999zz}.
However, the different models given in Ref. \cite{Thompson:1994zz} showed that already a calculation with only 
\(s\)-wave \linine-\(n\) interactions can yield momentum distributions
in agreement with experimental data at an acceptable level at low momenta.
More recently, Casal and Moro achieved a reasonable description of the \linine\((d,p)\)\liten~reaction using \(1^-\)/\(2^-\) \(s\)-wave virtual states and \(1^+\)/\(2^+\) \(p\)-wave resonances around 500 keV~\cite{Moro:2019whw}.
There seems to be no conclusive evidence for a \(p\)-wave resonance in the \linine-\(n\) system significantly below 500 keV, despite many investigations of this system over the years \cite{Kondev:2021lzi,Zinser:1997da,Bohlen:1997,Jeppesen:2006xcn,Simon:2007rtt}.
We therefore follow the Halo EFT power counting, which stipulates that \(p\)-wave resonances at energies $\approx 500$ keV, i.e., corresponding to momenta of order $M_{halo}$, produce only a NLO effect in the E1 response, unless they are kinematically enhanced because the experimental energy is tuned to the resonance energy. 

The paper is structured as follows.
In Sec.~\ref{sec:11Li} we derive the leading-order wave function of ${}^{11}$Li in Halo EFT.
We write down the Faddeev equations for this system and define the \(s\)-wave interactions that govern its structure
at leading order in Halo EFT. We also elucidate the differences in these equations that result
because ${}^9$Li is a spin-$3/2$ core and not a spin-0 core and discuss the circumstances under
which the more complex spin situation encountered in this problem reduces to the case of a spin-$0$ core.
In Sec.~\ref{sec:CDwithFSI} we present our calculations of the $E1$ matrix element of the ${}^{11}$Li
ground state and the non-interacting $nnc$ scattering state and show how to use M{\o}ller operators to incorporate
$nc$ and $nn$ final-state interactions. 
Moreover, we discuss the influence of different-spin \(nc\) interaction channels.
In Sec.~\ref{sec:higherorder} we show how to develop
approximations to the final-state scattering wave function that preserve the non-energy-weighted sum
rule, before concluding in Sec.~\ref{sec:conclusion}.


\section{\texorpdfstring{\(\boldsymbol{^{11}}\)Li in Halo EFT}{Li-11 in Halo EFT}}
\label{sec:11Li}

A convenient ingredient for describing \lieleven~and calculating observables is its wave function.
In our Halo EFT description this is the wave function of a three-body system.
The wave function is a concept commonly known from quantum mechanics, that can also appear in a nonrelativistic field theory through its relation to the vertex function that is the residue of the three-body scattering amplitude at the bound-state pole  (see, e.g., the  review \cite{Hammer:2017tjm}).
Since the corresponding half-off-shell amplitude appears in many calculations the wave function is a useful intermediate step 
in computations of observables, and 
can be seen as a way to modularize the calculations.

We will calculate the wave function from Faddeev amplitudes determined by Faddeev equations.
Our approach for the ground state is similar to the one of Canham and Hammer in Ref.~\cite{Canham:2008jd}, where two-neutron halos were described as three-body systems
with \(s\)-wave interactions in Halo EFT. However, we go beyond the treatment of Ref.~\cite{Canham:2008jd} since we do not assume that the spin of ${}^9$Li and the total angular momentum of the ${}^{11}$Li bound state are both zero. Instead we analyze the spin structure of \(^{11}\)Li in detail.

\subsection{Jacobi momenta and Faddeev equations}

Before discussing the Faddeev equations through which we calculate the ground state wave function and introducing the different interactions therein,
we first summarize how the different momenta in the three-body system can be described.
Typically Jacobi momenta are employed, whereby the system is described in terms of a relative
momentum within a two-body subsystem and the momentum between the third particle and the subsystem.
The third particle is called the spectator. Since there are three different choices of spectator possible there are three different Jacobi co-ordinate systems and these 
are labeled by the particle chosen as spectator. 
The definition of the Jacobi momenta \(\v{p}_i\) and \(\v{q}_i\) with respect to spectator \(i\) in a system with masses $\{m_i,m_j,m_k\}$ and
momenta $\{\v{k}_i\,\v{k}_j,\v{k}_k\}$ reads
\begin{align}
    \v{p}_i &\coloneqq \mu_{jk}      \K{ \frac{\v{k}_j}{m_j} - \frac{\v{k}_k}{m_k}              } \,, &
    \v{q}_i &\coloneqq \mu_{i\K{jk}} \K{ \frac{\v{k}_i}{m_i} - \frac{\v{k}_j + \v{k}_k}{M_{jk}}}  \, .
\end{align}

The Faddeev equations for the abstract Faddeev components \(\ket{F_i}\) can be written as
\begin{equation}
    \ket{F_i} = \sum_{j \neq i} G_0 t_j \ket{F_j} \,,
\end{equation}
where \(G_0\) is the free Green's function and \(t_j\) is the two-body t-matrix for the $(ik)$ sub-system, embedded in the three-body Hilbert space.
The concrete expression in the case of a two-neutron system with one \(nn\) and one \(nc\) interaction reads
\begin{align}\label{eq:fd_f_c}
    F_c{\K{q}} &= \K{ 1 + (-1)^{l(\zeta_c) + s(\xi_c)} } \rint{\qp} X_{cn}{\K{q,\qp}} 4\pi tau_n{\K{q'}} \ibraket{c}{\xi_c}{\xi_n}{n} F_n{\K{\qp}} \,, \\ 
    F_n{\K{q}} &= \ibraket{n}{\xi_n}{\xi_c}{c} \rint{\qp} X_{nc}{\K{q,q'}} 4\pi \tau_{c}{\K{\qp}} F_c{\K{\qp}}
    - \imel{n}{\xi_n}{\pmospin}{\xi_n}{n} \rint{\qp} X_{nn}{\K{q,q'}} 4\pi \tau_n{\K{\qp}} F_n{\K{\qp}} \,. \label{eq:fd_f_n} 
\end{align}
whereby the functions \(F_i{\K{q}}\) are related to the abstract components \(\ket{F_i}\) via
\(F_i{\K{q}} \coloneqq \rint{\p} g_{l(\zeta_i)}{\K{p}} \ibraket{i}{p,q;\zeta_i}{F_i}{}\) with some orbital angular momentum quantum numbers \(\zeta_i\).
The regulators are given by the \(g_l\).
The \(X_{ij}\) are the so-called ``kernel functions'' originating from the evaluation of free Green's functions between states
differing in the spectator. The expressions can be found in Ref. \cite{Hammer:2017tjm}\footnote{
    Note that we use the definition
    \(X_{ij}{\K{q,q'}} \coloneqq \rint{\p} \rint{\pp} g_{l(\zeta_i)}{\K{\p}} g_{l(\zeta_j)}{\K{\pp}} \imel{i}{p,q; \zeta_i}{G_0}{\pp,\qp; \zeta_j}{j} \).
    In the case of sharp-cutoff regularization via the \(g_l\) these can be neglected at low momenta.
    If additionally the already mentioned interaction channels are \(s\)-wave, one can use
    the expressions from Ref. \cite{Hammer:2017tjm}. The notation is slightly different, whereby the relation 
    \(X_{nc}{\K{q, q'}} = X_{00}^{n}{\K{q,q';B_3}}\) holds.
    Moreover, the relation \(X_{cn}{\K{q, q'}} = X_{nc}{\K{q', q}}\) can be employed.
    The function \(X_{nn}\) has a \(\pmospatial\) in front of the \(G_0\).
    Here the relation \(X_{nn}{\K{q, q'}} = X_{00}^{c}{\K{q,q';B_3}}\) can be used.
    Alternatively, the regulator effects on the kernel functions could be explicitly taken into account by evaluating
    some of the integrals in the functions numerically.
    This is discussed for \({}^{6}\)He in Ref. \cite{Gobel:2019jba}. 
}.
The functions \(\tau_i\) are related to t-matrix elements and will be defined in \cref{eq:def_tau_i}.
A three-body force can be included in these Faddeev equations by replacing \(X_{nn}{\K{q,q'}}\) by \(X_{nn}{\K{q,q'}} + h\) with
\(h\) being some three-body force parameter, see, e.g., Ref. \cite{Hammer:2017tjm}.
Here the multiindex \(\xi_n\) specifies the spin state of the three-body system seen from the neutron as spectator
when the \(nc\) subsystem is in the spin state of the \(nc\) interaction channel.
Analogously, \(\xi_c\) specifies the spin state seen from the core as spectator when the \(nn\) subsystem is the 
the spin state of the \(nn\) interaction channel\footnote{
    Note the semantic difference between ``specifies the spin state'' and ``is the spin state'': e.g., \(\xi_c\) is just a
    collection of quantum numbers, denoted by an subscript \(c\). If applied with the core as spectator, then this collection specifies the spin state of the \(nn\) interaction channel.
    I.e., \(\iket{\xi_c}{c}\) is the spin state of the \(nn\) interaction channel.
    Nevertheless \(\xi_c\) is just a collection of quantum numbers and can be also applied by using another spectator, which results
    in a different spin state.
    This means that \(\iket{\xi_c}{n}\) is a mathematically valid expression.
    However, it is not necessarily an allowed spin state of this system
    (e.g., \(\iket{\xi_c}{n}\) would require neutrons of spin \(3/2\)).
}.
In the case of a spinless core these overlaps read
\begin{align}
    \ibraket{n}{\xi_n}{\xi_c}{c} &= -1 \,, \\
    \imel{n}{\xi_n}{\pmospin}{\xi_n}{n} &= -1 \,, 
\end{align}
and one obtains Faddeev equations equivalent to the ones from Ref. \cite{Canham:2008jd}: the equations are the same apart from a relative minus sign in the definition of
\(F_c\) from the \(F_n\) between our version and the one from Ref. \cite{Canham:2008jd}.
If one continues this comparison to the level of wave functions,
one finds that the total wave functions are equivalent up to overall
minus signs that depend on the spectator and are not observable: in the case of \(\Psi_c{\K{p, q}}\) 
there is a relative minus sign, while in the case of \(\Psi_n{\K{p,q}}\) there is no sign difference.

\subsection{\texorpdfstring{Spin structure of the interactions and of \(\boldsymbol{^{11}}\)Li}{Spin structure of the interactions and of Li-11}}

Now that we have seen the Faddeev equations and the way they are influenced by spin states, we want to discuss the spin structure of the interactions specified as t-matrices as well as the overall
spin structure of the two-neutron halo in detail.
\lieleven~and \linine~have the same non-zero overall angular momentum: \(J = s_c = 3/2\).
This makes the $nc$ dynamics more complicated, since the neutron and ${}^9$Li can interact in either the spin-1 or spin-2 channel.

We treat the core spin in two different ways:
\begin{itemize}
\item In  \cref{ap:equiv_one_channel_two_channels}, we show that if the $nc$ interactions in both the spin-1 and spin-2 channels have the same strength then the three-body Hamiltonian can be separated into two terms. In one
the \(nn\) system is in a spin-0 configuration and in the other  the two neutrons form a spin-1 pair. The interactions in the first Hamiltonian are the same as those in the Hamlitonian that treats the ${}^9$Li and ${}^{11}$Li as spin-0 particles. Therefore in this case we can just carry out a calculation that neglects the core spin, since that quantum number does not play a role in the dynamics of the system. 
For the ground state we explicitly checked the equivalence of the calculation having two \(nc\) spin channels of equal interaction strength
with the spinless calculation\footnote{
    Note that we will sometimes call this calculation ``spinless'' for simplicity.
    However, this adjective refers only to the core and the overall halo nucleus.
    The spins of the neutrons are always included.
} by verifying that the numerical results for the wave functions were the same. 
Note that this equivalence statement only refers to the spatial parts.
If one wants to assemble the full state, the spatial solution has to be combined with the correct spin state.

\item In the other approach, we take the core spin into account and assume that the leading-order \(nc\) interaction  is only in the \(s_c + 1/2\) (spin-2) channel. The interaction in the \(s_c - 1/2\) (spin-1) channel is taken to be subleading.
\end{itemize}

Previous Halo EFT treatments basically used the first approach, whereby \cref{ap:equiv_one_channel_two_channels} can be seen as a formalization
of an argument given in Ref. \cite{Hammer:2017tjm}.

We now describe the spin configurations.
We focus on the case with the \(nc\) interaction in the \(s_c+1/2\) channel and mention the simplifications when the case with two equal \(nc\) 
interactions is realized through a calculation taking $s_c=0$.
For specifying the interaction channels in the three-body system we use a basis of states of definite \(L\) and \(S\) but in general indefinite \(J\).
The interaction channel for the \(nn\) interaction, which is given by the conditions
\begin{equation}
    l = 0\,, \quad s = 0 \quad \textrm{(seen from core)}\,,
\end{equation}
can be written as
\begin{equation}
    \iket{\K{0,\lambda}\lambda,\mu}{c} \iket{\K{0,\frac{3}{2}}\frac{3}{2},M_S}{c}
\end{equation}
where the core is used as the spectator.
Here we have specified the states in \(LS\)-coupling, using the following notation:
\begin{equation}
    \iket{\K{l,\lambda}L,M_L}{i} \iket{\K{s,\sigma}S,M_S}{i} \,,
\end{equation}
where the total orbital angular momentum of the three-body system $\v{L}=\v{l} + \v{\lambda}$, with $\v{l}$ the orbital angular momentum of the $jk$ subsystem and $\v{\lambda}$ the orbital angular momentum of particle \(i\) relative to the $jk$ pair. Similarly the total spin $\v{S}$ is composed of the spin of the pair plus the spin of the spectator: $\v{S}=\v{s} + \v{\sigma}$. In the case where the core spin does play a role the \(nc\) pair interacts when
\begin{equation}
    l = 0\,, \quad s = 2 \quad \textrm{(seen from the spectator neutron)},
\end{equation}
which can be written as
\begin{equation}
    \iket{\K{0,\lambda}\lambda,\mu}{n} \iket{\K{2,\frac{1}{2}}\frac{3}{2},M_S}{n},
\end{equation}
where the neutron is used as spectator. Since for the ground-state calculation we restrict ourselves to \(L=0\) 
it follows that \(S=3/2\) has to hold so that the ${}^{11}$Li ground state has the correct angular momentum. The spins
are not affected by the \(E1\) operator, but, in contrast, the \(L\) can change, which is why we want to consider both \(L=0\) (bound state)
and \(L=1\) (scattering state created after action of electric dipole operator). 

In the calculations where the core spin is neglected all the \(3/2\) have to be replaced by zeros and \(M=0\) holds. This makes 
the transformation of the spin states between the different spectators simple:
\begin{equation}
    \iket{\K{0,0}0,0}{c} = - \iket{\K{\oh,\oh}0,0}{n} \,.
\end{equation}

Now, if the core spin is taken into account the corresponding relation reads
\begin{equation}\label{eq:recoupling_pw_gs}
    \iket{\K{0,\frac{3}{2}}\frac{3}{2},M_S}{c} = - \frac{\sqrt{5/2}}{2} \iket{\K{2,\frac{1}{2}}\frac{3}{2},M_S}{n} - \frac{\sqrt{3/2}}{2} \iket{\K{1,\frac{1}{2}}\frac{3}{2},M_S}{n} 
\end{equation}
If we write the spins in the subsystems out (in square brackets), the relation reads
\begin{equation}
    \iket{\K{\Ke{\frac{1}{2},\frac{1}{2}}0,\frac{3}{2}}\frac{3}{2},M_S}{c} = - \frac{\sqrt{5/2}}{2} \iket{\K{\Ke{\frac{1}{2},\frac{3}{2}}2,\frac{1}{2}}\frac{3}{2},M_S}{n} - \frac{\sqrt{3/2}}{2} \iket{\K{\Ke{\frac{1}{2},\frac{3}{2}}1,\frac{1}{2}}\frac{3}{2},M_S}{n} 
\end{equation}

As an approximation we take only the following partial-wave component of the \lieleven~ground state into account
\begin{equation}\label{eq:wf_pw}
    \iket{\K{0,0}0\K{0,\frac{3}{2}}\frac{3}{2};\frac{3}{2},M}{c} = \iket{\K{0,0}0,0}{c} \iket{\K{0,\frac{3}{2}}\frac{3}{2},M}{c} \,,
\end{equation}
which was easily recoupled from \(jJ\)-coupling into \(LS\)-coupling.
In \(jJ\)-coupling we use the following notation:
\begin{equation}
    \iket{\K{l,s}j \K{\lambda, \sigma} J; J, M}{i} \,.
\end{equation}
Again, the relation for the spinless case is obtained by replacing the \(3/2\) in \cref{eq:wf_pw} by zeros.

In order to refer to certain partial-wave states compactly, we use multiindices. We use the following naming convention here:
multiindices specifying a full state in \(jJ\)-coupling are denoted by \(\Xi\), multiindices denoting the pure spatial part are denoted
by \(\zeta\), and those denoting the spin part are denoted by \(\xi\).
We introduce the following abbreviations:
\begin{align}
    \Xi_c^{(M)} &\coloneqq \K{0,0}0\K{0,\frac{3}{2}}\frac{3}{2};\frac{3}{2},M \,,\label{eq:multiindexXi}\\
    \zeta_c     &\coloneqq \K{0,0}0,0 \,,\\
    \xi_c^{(M)} &\coloneqq \K{0,\frac{3}{2}}\frac{3}{2},M \,.
\end{align}
Using these \cref{eq:wf_pw} can be written as
\begin{equation}
    \iket{\Xi_c^{(M)}}{c} = \iket{\zeta_c \vphantom{\xi_c^{(M)}} }{c} \iket{\xi_c^{(M)}}{c} \,.
\end{equation}

\subsection{Spatial structure of the interactions}

Now that we have discussed the spin structure of the interactions as well as of \(^{11}\)Li,
we have to discuss the general nature of the interactions.
For our EFT calculation it is particularly useful to use the interactions in the form of t-matrices,
as in their denominators the different order terms can be identified.
The leading-order reduced t-matrix of the \(nn\) interaction reads
\begin{equation}
    \tau_{nn}{\K{E}} = \frac{1}{4\pi^2 \mu_{nn}} \frac{1}{1/a_{nn} + \ci k} \,,
\end{equation}
whereby \(a_{nn}\) is the \(nn\) scattering length and the relation \(k = \sqrt{2\mu_{ij} E}\) holds.
The relation between the reduced t-matrix matrix element and the full two-body t-matrix element is
\begin{equation}
    \mel{p,l,s}{t_{ij}{\K{E}}}{p', l',s'} = 4\pi \kd{l}{l'} \kd{s}{s'} \kd{l_{ij}}{l} \kd{s_{ij}}{s} g_{l_{ij}}{\K{p}} \tau_{ij}{\K{E}} g_{l_{ij}}{\K{p'}} \,.
\end{equation}
The \(nc\) interaction is given at leading order by the \(s\)-wave virtual state\footnote{
    Note that the atomic mass evaluation \cite{Wang:2021xhn,Huang:2021nwk} characterizes this state as an \(s\)-wave resonance.
    In principle the effective nuclear \linine-\(n\) \(s\)-wave potential could have a barrier that produces a low-energy resonance.
    However, the main references given by the atomic evaluation in Ref. \cite{Huang:2021nwk}
    characterize this state as a virtual state \cite{Zinser:1994ii,Aksyutina:2008zz}.
    We follow the original references. 
} characterized by the 
virtual state energy.
The t-matrix reads
\begin{equation}
    \tau_{nc}{\K{E}} = \frac{1}{4\pi^2 \mu_{nc}} \frac{1}{\gamma_{nc} + \ci k} \,.
\end{equation}
The virtual state momentum \(\gamma_{nc}\) is related to the virtual state energy \(E_{nc}\) according to
\(\gamma_{nc}=-\sqrt{2\mu_{nc} E_{nc}}\).

The embedding of the t-matrices in the three-body space is given by
\begin{equation}
    \imel{i}{p,q;\zeta,\xi}{t_i{\K{E_3}}}{p',q';\zeta',\xi'}{i} = \kd{\zeta}{\zeta'} \kd{\xi}{\xi'} 
        \mel{p,l{\K{\zeta}},s{\K{\xi}}}{ t_{jk}{\K{E_3 - \frac{q^2}{2\mu_{i(jk)}} }} }{{p',l{\K{\zeta'}},s{\K{\xi'}}}} \frac{\delta{\K{q-q'}}}{q'^2} \,.
\end{equation}
This motivates the introduction of
\begin{equation}\label{eq:def_tau_i}
    \tau_i{\K{q;E_3}} \coloneqq \tau_{jk}{\K{ E_3 - \frac{q^2}{2\mu_{i(jk)}} }} \,,
\end{equation}
for compact notation.
The three-body energy \(E_3\) is in our case given by \(-B_3 = -S_{2n}\).
Sometimes this second argument of \(\tau_i\) is omitted.

\subsection{From the Faddeev amplitudes to wave functions}

Now that we have discussed the Faddeev equations and the effective interactions that appear in them, we turn our attention to the 
wave function and how it is obtained from the Faddeev amplitudes.
The starting point is the relation between the abstract Faddeev amplitudes \(\ket{F_i}\) and the overall state \(\ket{\Psi}\).
It is given by
\begin{equation}
    \ket{\Psi} = \sum_i G_0 t_i \ket{F_i} \,.
\end{equation}
We can now define individual Faddeev components of the wave function that appear here as \(\psi_i{\K{p,q}} = G_0^{(i)}{\K{p,q;E_3}} \tau_i{\K{q;E_3}} F_i{\K{q}}\).

By projecting on a reference state and using the representations of the Faddeev amplitudes one obtains the overall wave function
in a particular partial wave specified in terms of a spatial multiindex \(\zeta\) and a spin multiindex \(\xi\) as:
\begin{align}\label{eq:Psi_c_arb_pw}
    \Psi_{c;\zeta,\xi}{\K{p,q}} &= \sum_{M'} \psi_c{\K{p,q}} \kd{\zeta}{\zeta_c} \kd{\xi}{\xi_c^{(M')}} + \sum_{M'} \K{1+\K{-1}^{l-s}} \kd{L}{0} \kd{M_L}{0} \kd{\lambda}{l} \frac{\sqrt{2l+1}}{2} \K{-1}^l \nonumber \\
    &\quad \times \ibraket{c}{\xi}{\xi_n^{(M')}}{n} \int_{-1}^1 \dd{x} P_l{\K{x}} \psi_n{\K{\kcnp{\K{p,q,x}},\kcnq{\K{p,q,x}}}}\,.
\end{align}

It turns out that in our case
\begin{align}
    \Psi_c{\K{p,q}} \coloneqq \ibraket{c}{p,q;\Xi_c^{(M)}}{\Psi}{} = \ibraket{c}{p,q;\zeta_c, \xi_c^{(M)}}{\Psi}{} \, ,
\end{align}
with the multi-index defined as in Eqs.~(\ref{eq:multiindexXi}),
is the most important partial-wave component of the wave function.
Moreover, the quantum number \(M\) is undetermined, as the results are independent of \(M\). (The superscript \(M\) of the multiindex \(\Xi_c\) is therefore omitted hereafter for brevity.)
The expression obtained from \cref{eq:Psi_c_arb_pw} for this piece of the wave function is:
\begin{align}
    \Psi_{c}{\K{p,q}} = \psi_c{\K{p,q}} + \ibraket{c}{\xi_c}{\xi_n}{n} \int_{-1}^1 \dd{x} P_0{\K{x}} 
        \psi_n{\K{\kcnp{\K{p,q,x}},\kcnq{\K{p,q,x}}}} \label{eq:Psi_c} \,.
\end{align}
This is the piece of the wave function considered by Canham and Hammer in Ref.~\cite{Canham:2008jd}. 

We also checked the importance of wave function components corresponding to higher angular momenta for both the $nn$ pair and the core relative to that pair, i.e., 
\begin{equation}\label{eq:Psi_c_l}
    \Psi_c^{(l)}{\K{p,q}} \coloneqq \ibraket{c}{p,q;\zeta_c^{(l)}, \xi_c}{\Psi}{} \,,
\end{equation}
with the multiindex  \( \zeta_c^{(l)} \coloneqq \K{l, l} 0, 0 \,\).
Our calculations show that \(\Psi_c^{(2)}\) is typically suppressed by a factor of 100 or more compared to \(\Psi_c\)  in terms of their respective maxima. 
This means that here, as  was done in Ref.~\cite{Canham:2008jd}, we will use only the \(l = 0\) states as reference states.

Moreover, all these equations are under the assumption that only \(L=0\) states appear in the bound-state wave function.
Because the core has spin 3/2 it is in principle possible for \(L=2\) components to also be present in the \(J=3/2^-\)
\lieleven~ground state. However, the leading-order three-body force in Halo EFT is operative only for \(L=0\):
three-body forces that mix angular momenta or are operative in other \(L\) channels occur in Halo EFT, but only at higher orders. Therefore Halo EFT predicts a \lieleven~state with only \(L=0\) (at leading order). In the case of \(L = 0\) and only \(s\)-wave interactions it is no loss of generality to
assume that \(t_i\) projects also on \(L = 0\) and on \(\lambda = 0\).
In the spin space, we project not only in \(s\) but also in \(\sigma\) with no loss of generality.

Before presenting our results, we briefly discuss the parameters
and renormalization conditions applied in
our calculation.
The \linine-n virtual state, which nature was already discussed, is characterized by an energy of \(E_{nc} = \) 26(13) keV \cite{Wang:2021xhn}.
We use it to calculate a virtual-state momentum \(\gamma_{nc} = -\sqrt{2\mu_{nc} E_{nc}}\). This parameterizes the \(nc\) interaction.
The corresponding scattering length \(a_{nc}\) amounts to -29.8 fm.
The \(nn\) interaction is parameterized by \(a_{nn} = \) -18.7 fm \cite{GonzalezTrotter:2006wz}.
Meanwhile, we use the two-neutron separation energy
\(S_{2n} = \) 0.369 MeV \cite{Wang:2021xhn} to renormalize the three-body energy.
The mass of the \linine~core is approximated by \(A m_n\) with the neutron mass \(m_n\) and \(A=9\).
The two-body systems as well as the three-body system are regulated using sharp cutoffs.
Moreover, the three-body cutoff \(\Lambda\) is chosen to be equal to the two-body cutoff, \(\Lambda =\) 400 MeV.
In order to check the convergence of the results, we compare to calculations with \(\Lambda = \) 300 MeV and
two thirds of the mesh points for discretizations and integrations.

\subsection{\texorpdfstring{Radius of \(\boldsymbol{{}^{11}}\)Li}{Radius of Li-11}}
\label{sec:radius}

The matter radius \(r_c\) which is the distance between the core and the halo's center of mass can be extracted as root-mean-square (rms) radius
from the so-called form factor \(\mathcal{F}_c\) via
\begin{equation}
    \expval{r_c^2} = -6 \frac{\dd{\mathcal{F}_c{\K{k^2}}}}{\dd{k^2}} \bigg|_{k^2 = 0} \K{\frac{2}{A+2}}^2 \,.
\end{equation}
The factor \(2/(A+2)\) stems from the conversion between the distance \(y_c\), which corresponds to the momentum \(q_c\), and 
\(r_c\).
The expression for the form factor reads \cite{Hammer:2017tjm} 
\begin{equation}
    \mathcal{F}_c{\K{k^2}} = \drint{\p}{\q} \Psi_c^*{\K{p, q}} \Psi_c{\K{p, q+k}} \,.
\end{equation}
We use only the \(l=0\) component of the wave function
in our calculations, since, as discussed above, other components are suppressed by
at least a factor of 100.
This way we obtain \(\sqrt{\expval{r_c^2}} = 0.87\) fm with an numerical uncertainty of roughly 0.02 fm and LO EFT uncertainty
of approximately \(\sqrt{\expval{r_c^2}} \sqrt{S_{2n}/E^*} \approx \) 0.32 fm.
Hereby, the two-neutron separation energy of \lieleven~is given by \(S_{2n}\) and the excitation energy of \linine~is given
by \(E^*\).
In order to compare our value with experimental data we use the experimental values for the rms charge radii of \linine~and \lieleven,
\(\sqrt{\expval{r_9^2}}\) and \(\sqrt{\expval{r_{11}}}\) from
which $\expval{r_c^2}$ can be obtained.
These can be obtained from isotope shift measurements.
The first values were obtained in Ref. \cite{Sanchez:2006zz}, while
we use the more current ones from Ref. \cite{Nortershauser:2011zz}.
These yield \(\sqrt{\expval{r_c^2}} = \sqrt{\expval{r_{11}^2} - \expval{r_{9}^2}} = 1.04 \pm 0.14\) fm.
If we also include the mean-square neutron charge radius, \(\expval{r_n^2}=-0.1161 \pm 0.0022\) fm\(^2\) \cite{ParticleDataGroup:2020ssz},
we obtain \(\sqrt{\expval{r_c^2}} =1.08 \pm 0.14\) fm by using the formula from Ref. \cite{Horiuchi:2007ww}.
Our theoretical result is in good agreement with both values.

Furthermore,  it is interesting to use our \(\sqrt{\expval{r_c^2}}\) together with the \(\sqrt{\expval{r_9^2}}\) from experiment to calculate
an \(\sqrt{\expval{r_{11}^2}}\).
We obtain \(\sqrt{\expval{r_{11}^2}} = \sqrt{ \expval{r_c^2} + \expval{r_9^2}} = 2.41 \pm 0.13\) fm (including the rms neutron charge radius changes the result
by less than 0.02 fm).
Our value is not far from the experimental result of \(2.48 \pm 0.04\) fm
and agrees within uncertainties.
This means that a LO EFT three-body description of \lieleven~is able to describe the charge radius without explicitly including
core excitation.
It will be interesting to see if this persists at NLO.
Finally, we want to mention that \(r_c\) is related to the neutron-pair-to-core distance \(r_{c(nn)}\) by
\(11\,r_c/2 = r_{c(nn)}\).
Thereby we obtain for
\(\langle r_{c(nn)}^2 \rangle^{1/2}\)
a value of 4.8 fm with an LO EFT uncertainty of 1.8 fm.
This large \(nn-c\) distance is another strong manifestation of the halo structure of \lieleven.


\section{E1 Coulomb dissociation}

\label{sec:CDwithFSI}

In this section, we investigate the \(E1\) strength function without final-state interactions (FSIs)
as well as the impact that $nn$ and $nc$ FSI separately have on this strength function. Before showing and discussing the results, we give a diagrammatic overview of these calculations
in terms of the Feynman diagrams for the matrix elements of the \(E1\) operator.
The diagrams are shown in \cref{fig:e1_diagrams}.
Final-state interactions are those interactions happening after the \(E1\) breakup of the halo nucleus. 
(In this section, only one FSI will be included at a time. The inclusion of
multiple interactions at once will be discussed in the following section.)


\newcommand{\lbox}[4]{
  \filldraw[fill=#3, draw=#4,thick] (#1) rectangle (#2);
}

\newcommand{\ellipse}[4]{

  \coordinate (M) at ($0.5*(#1)+0.5*(#2)$);
  \filldraw[fill=#3, draw=black,thick] let \p1=($(#1)-(#2)$) in (M) ellipse (\x1 and 0.5*\y1);
  \node at (M) {#4};
}

\newcommand{\thrlines}[1]{
  \coordinate (L2) at (#1);
  \coordinate (L1) at ($(L2) + (0,\mh)$);
  \coordinate (L3) at ($(L2) - (0,\mh)$);
  \coordinate (R1) at ($(L1) + 0.7*(\ml,0)$);
  \coordinate (R2) at ($(L2) + 0.7*(\ml,0)$);
  \coordinate (R3) at ($(L3) + 0.7*(\ml,0)$);
  \draw[thick,\dnc] (L1) -- (R1);
  \draw[thick,\dnc] (L2) -- (R2);
  \draw[thick,dashed,\dcc] (L3) -- (R3);
}

\newcommand{\fvertex}[8]{
 \coordinate (I1) at ($(#1)+0.2*(#4)-0.2*(#1)$);
 \coordinate (I2) at ($(#2)+0.2*(#4)-0.2*(#1)$);
 \coordinate (I3) at ($(#3)-0.2*(#4)+0.2*(#1)$);
 \coordinate (I4) at ($(#4)-0.2*(#4)+0.2*(#1)$);
 \coordinate (M) at ($0.25*(#1)+0.25*(#2)+0.25*(#3)+0.25*(#4)$);

 \draw[black,thick,#7] (#1) -- (I1);
 \draw[black,thick,#7] (#4) -- (I4);
 \draw[black,thick,#6] (#2) -- (I2);
 \draw[black,thick,#6] (#3) -- (I3);

 \draw[black,thick,#7] (I4) -- (I1);
 \draw[black,thick,#6] (I3) -- (I2);

 \filldraw[fill=#5, draw=#8,thick] let \p1=($(#2)-(#1)$) in (M) ellipse (0.5*\y1 and 0.5*\y1);
}

\newcommand{\cintthb}[4]{
  \coordinate (M) at (#1);
  \coordinate (HS) at (#2,0);
  \coordinate (VS) at (0,#2);
  \coordinate (C1) at ($(M)-(HS)-(VS)$);
  \coordinate (C2) at ($(M)-(HS)+(VS)$);
  \coordinate (C2U) at ($(C2)+0.07*(HS)+0.07*(VS)$);
  \coordinate (C2L) at ($(C2)-0.07*(HS)-0.07*(VS)$);
  \coordinate (C3) at ($(M)+(HS)+(VS)$);
  \coordinate (C3U) at ($(C3)-0.07*(HS)+0.07*(VS)$);
  \coordinate (C3L) at ($(C3)+0.07*(HS)-0.07*(VS)$);
  \coordinate (C4) at ($(M)+(HS)-(VS)$);
  \coordinate (MRU) at ($(M)-0.07*(HS)+0.07*(VS)$);
  \coordinate (MRL) at ($(M)+0.07*(HS)-0.07*(VS)$);
  \coordinate (MLU) at ($(M)+0.07*(HS)+0.07*(VS)$);
  \coordinate (MLL) at ($(M)-0.07*(HS)-0.07*(VS)$);
  \draw[black,thick,dashed,#4] (C1) -- (M);
  \draw[black,thick,dashed,#4] (C4) -- (M);
  \draw[black,thick,#3] (C2U) -- (MLU);
  \draw[black,thick,#3] (C2L) -- (MLL);
  \draw[black,thick,#3] (C3U) -- (MRU);
  \draw[black,thick,#3] (C3L) -- (MRL);
  \filldraw[fill=\dtbv, draw=\dtbv,thick] (M) ellipse (0.12 and 0.12);
}

\newcommand{\cint}[5]{
  \coordinate (M) at (#1);
  \coordinate (HS) at (#2,0);
  \coordinate (VS) at (0,#2);
  \coordinate (C1) at ($(M)-(HS)-(VS)$);
  \coordinate (C2) at ($(M)-(HS)+(VS)$);
  \coordinate (C3) at ($(M)+(HS)+(VS)$);
  \coordinate (C4) at ($(M)+(HS)-(VS)$);
  \draw[thick,#3] (C1) -- (M);
  \draw[thick,#3] (C4) -- (M);
  \draw[thick,#4] (C2) -- (M);
  \draw[thick,#4] (C3) -- (M);
  \filldraw[fill=#5, draw=#5,thick] (M) ellipse (0.12 and 0.12);
}

\newcommand{\cintd}[7]{
  \coordinate (M) at (#1);
  \coordinate (HS) at (#2,0);
  \coordinate (VS) at (0,#2);
  \coordinate (C1) at ($(M)-(HS)-(VS)$);
  \coordinate (C2) at ($(M)-(HS)+(VS)$);
  \coordinate (C3) at ($(M)+(HS)$);
  \coordinate (C3U) at ($(C3)+0.1*(VS)$);
  \coordinate (C3L) at ($(C3)-0.1*(VS)$);
  \coordinate (MU) at ($(M)+0.1*(VS)$);
  \coordinate (ML) at ($(M)-0.1*(VS)$);
  \draw[thick,#3,#6] (C1) -- (ML);
  \draw[thick,#3,#5] (C3L) -- (ML);
  \draw[thick,#4,#7] (C2) -- (MU);
  \draw[thick,#4,#5] (C3U) -- (MU);
  \filldraw[fill=#5, draw=#5,thick] (M) ellipse (0.12 and 0.12);
}

\newcommand{\dline}[5]{
  \coordinate (L) at (#1);
  \coordinate (R) at (#2);
  \coordinate (VS) at (0,#3);
  \coordinate (LU) at ($(L)+0.1*(VS)$);
  \coordinate (LL) at ($(L)-0.1*(VS)$);
  \coordinate (RU) at ($(R)+0.1*(VS)$);
  \coordinate (RL) at ($(R)-0.1*(VS)$);
  \draw[black,thick,#4] (LL) -- (RL);
  \draw[black,thick,#5] (LU) -- (RU);
}

\newcommand{\ATdiagr}[9]{
  \coordinate (C1) at (0,0);
  \coordinate (C2) at (0,2*\mh);
  \coordinate (C3) at (\ml,2*\mh);
  \coordinate (C4) at (\ml,0);
  \fvertex{C1}{C2}{C3}{C4}{#4}{#2}{#3}{#8};
  \coordinate (C5) at (0,\maxh);
  \coordinate (C6) at (1.5*\ml,\maxh);
  \draw[black,thick,#1] (C5) -- (C6);
  \coordinate (C11) at (1.5*\ml,\maxh-2*\mh);
  \coordinate (C12) at (1.5*\ml,\maxh);
  \coordinate (C13) at (2.8*\ml,\maxh);
  \coordinate (C14) at (2.8*\ml,\maxh-2*\mh);
  \draw[black,thick,#2] (C3) -- (C11);
  \fvertex{C11}{C12}{C13}{C14}{#5}{#1}{#2}{#9};
  \coordinate (C7) at (2.8*\ml,0);
  \draw[black,thick,#3] (C4) -- (C7);
  \coordinate (C20) at ($(C13) - (0.2*\ml,0)$);
  \coordinate (C21) at ($(C7) + (0.2*\ml,0)$);
  \coordinate (C31) at ($0.5*(C20) + 0.5*(C21)$);
  \thrlines{C31};
  \ellipse{C21}{C20}{#6}{#7};
}

\newcommand{\Adiagr}[9]{
  \coordinate (C1) at (0,0);
  \coordinate (C2) at (0,\mh);
  \coordinate (C3) at (\ml,\mh);
  \coordinate (C4) at (\ml,0);
  \lbox{C2}{C4}{#4}{#8};
  \coordinate (C5) at (0,\maxh);
  \coordinate (C6) at (1.5*\ml,\maxh);
  \draw[black,thick,#1] (C5) -- (C6);
  \coordinate (C10) at (1.5*\ml,\maxh-\mh);
  \coordinate (C12) at (1.5*\ml,\maxh);
  \coordinate (C13) at (2.5*\ml,\maxh);
  \coordinate (C14) at (2.5*\ml,\maxh-\mh);
  \draw[black,thick,#2] (C3) -- (C10);
  \lbox{C12}{C14}{#5}{#9};
  \coordinate (C7) at (2.5*\ml,0);
  \draw[black,thick,#3] (C4) -- (C7);
  \coordinate (C20) at ($(C13) - (0.2*\ml,0)$);
  \coordinate (C21) at ($(C7) + (0.2*\ml,0)$);
  \coordinate (C31) at ($0.5*(C20) + 0.5*(C21)$);
  \thrlines{C31};
  \ellipse{C21}{C20}{#6}{#7};
}

\newcommand{\Adiagl}[5]{
  \coordinate (C4) at (0,0);
  \coordinate (C10) at (0,\maxh-\mh);
  \coordinate (C12) at (0,\maxh);
  \coordinate (C13) at (\ml,\maxh);
  \coordinate (C14) at (\ml,\maxh-\mh);
  \lbox{C12}{C14}{#1}{#5};
  \coordinate (C7) at (\ml,0);
  \draw[thick,#2] (C4) -- (C7);
  \coordinate (C20) at ($(C13) - (0.2*\ml,0)$);
  \coordinate (C21) at ($(C7) + (0.2*\ml,0)$);
  \coordinate (C31) at ($0.5*(C20) + 0.5*(C21)$);
  \thrlines{C31};
  \ellipse{C21}{C20}{#3}{#4};
}

\newcommand{\ATdiagl}[7]{
  \coordinate (C4) at (0,0);
  \coordinate (C11) at (0,\maxh-2*\mh);
  \coordinate (C12) at (0,\maxh);
  \coordinate (C13) at (1.3*\ml,\maxh);
  \coordinate (C14) at (1.3*\ml,\maxh-2*\mh);
  \fvertex{C11}{C12}{C13}{C14}{#1}{#2}{#6}{#7};
  \coordinate (C7) at (1.3*\ml,0);
  \draw[black,thick,dashed,#3] (C4) -- (C7);
  \coordinate (C20) at ($(C13) - (0.2*\ml,0)$);
  \coordinate (C21) at ($(C7) + (0.2*\ml,0)$);
  \coordinate (C31) at ($0.5*(C20) + 0.5*(C21)$);
  \thrlines{C31};
  \ellipse{C21}{C20}{#4}{#5};
}

\newcommand{\plabel}[3]{
  \coordinate (R1) at (#1);
  \coordinate (R2) at (#2);
  \coordinate (LS) at (-0.1*\ml,0);
  \coordinate (M1) at ($(R1)+(LS)$);
  \coordinate (M2) at ($(R2)+(LS)$);
  \coordinate (L1) at ($(M1)+(LS)$);
  \coordinate (L2) at ($(M2)+(LS)$);
  \coordinate (MM) at ($0.5*(M1)+0.5*(M2)$);

  \draw[thick,gray] (L1) -- (R1);
  \draw[thick,gray] (L2) -- (R2);
  \draw[thick,gray] (M1) -- (M2);
  \draw[fill=white,draw=none] (MM) circle [radius=0.15*\ml];
  \node[text=gray] at (MM) {#3};
}

\newcommand{\pqlabel}[3]{
  \coordinate (LS)  at (-0.1*\ml,0);
  \coordinate (PR1) at ($(#1)+2*(LS)$);
  \coordinate (PR2) at ($(#2)+2*(LS)$);
  \coordinate (T)   at ($(#3)+2*(LS)$);
  \coordinate (PM1) at ($(PR1)+(LS)$);
  \coordinate (PM2) at ($(PR2)+(LS)$);
  \coordinate (PMM) at ($0.5*(PM1)+0.5*(PM2)$);

  \coordinate (QR1) at ($(PMM)+2*(LS)$);
  \coordinate (QR2) at ($(T)+3*(LS)$);
  \plabel{PR1}{PR2}{\(\v{p}\)}
  \plabel{QR1}{QR2}{\(\v{q}\)}
}

\newcommand{\WFA}[7]{
  \coordinate (C1) at (0,\maxh);
  \coordinate (C2) at (0,0.5*\maxh);
  \coordinate (C3) at (0.5*\ml,0.5*\maxh);
  \coordinate (C4) at (0,0);
  \pqlabel{C1}{C2}{C4};
  \coordinate (C10) at (\ml,\maxh-\mh);
  \coordinate (C12) at (\ml,\maxh);
  \coordinate (C13) at (2*\ml,\maxh);
  \coordinate (C14) at (2*\ml,\maxh-\mh);
  \lbox{C12}{C14}{#4}{#7};
  \coordinate (C7) at (2*\ml,0);
  \draw[black,thick,#1] (C1) -- (C12);
  \draw[black,thick,#2] (C2) -- (C3) -- (C10);
  \draw[black,thick,#3] (C4) -- (C7);
  \coordinate (C20) at ($(C13) - (0.2*\ml,0)$);
  \coordinate (C21) at ($(C7) + (0.2*\ml,0)$);
  \coordinate (C31) at ($0.5*(C20) + 0.5*(C21)$);
  \thrlines{C31};
  \ellipse{C21}{C20}{#5}{#6};
}

\newcommand{\WFAcr}[7]{
  \coordinate (C1) at (0,\maxh);
  \coordinate (C2) at (0,0.5*\maxh);
  \coordinate (C3) at (0.5*\ml,0.5*\maxh);
  \coordinate (C4) at (0,0);
  \coordinate (C5) at (0.5*\ml,0);
  \coordinate (C6) at (\ml,0);
  \pqlabel{C1}{C2}{C4};
  \coordinate (C10) at (\ml,\maxh-\mh);
  \coordinate (C12) at (\ml,\maxh);
  \coordinate (C13) at (2*\ml,\maxh);
  \coordinate (C14) at (2*\ml,\maxh-\mh);
  \lbox{C12}{C14}{#4}{#7};
  \coordinate (C7) at (2*\ml,0);
  \draw[black,thick,#1] (C1) -- (C12);
  \draw[black,thick,#2] (C2) -- (C3) -- (C6) -- (C7);
  \draw[black,thick,#3] (C4) -- (C5) -- (C10);
  \coordinate (C20) at ($(C13) - (0.2*\ml,0)$);
  \coordinate (C21) at ($(C7) + (0.2*\ml,0)$);
  \coordinate (C31) at ($0.5*(C20) + 0.5*(C21)$);
  \thrlines{C31};
  \ellipse{C21}{C20}{#5}{#6};
}

\newcommand{\drawphotonline}{
  \coordinate (CPh1) at (0,0);
  \coordinate (CPh2) at (0,-0.5*\maxh);
  \draw[black,thick,decorate,decoration=snake] (CPh1) -- (CPh2);
  \node at (CPh1) [circle,fill,inner sep=1.5] {};
}

\newcommand{\WFAPhMod}[7]{
  \coordinate (C1) at (-0.5*\ml, \maxh);
  \coordinate (C2) at (-0.5*\ml, \daspar*\maxh);
  \coordinate (C3) at (0.5*\ml,  \daspar*\maxh);
  \coordinate (C4) at (-0.5*\ml, 0);
  \pqlabel{C1}{C2}{C4};
  \coordinate (C10) at (\ml,\daspar*\maxh);
  \coordinate (C12) at (\ml,\maxh);
  \coordinate (C13) at (2.2*\ml,\maxh);
  \coordinate (C14) at (2.2*\ml,\daspar*\maxh);

  \draw[thick,#1] (C12) -- (C13);
  \draw[thick,#2] (C10) -- (C14);
  \coordinate (CM) at ($0.5*(C12)+0.5*(C14)-(0.2*\ml,0)$);
  \draw[fill=#4,draw=#7] (CM) circle ({0.5*(1-\daspar*\maxh)});
  
  \coordinate (C7) at (2.2*\ml,0);
  \draw[black,thick,#1] (C1) -- (C12);
  \draw[black,thick,#2] (C2) -- (C3) -- (C10);
  \draw[black,thick,#3] (C4) -- (C7);
  \coordinate (C20) at ($(C13) + (-0.1*\ml,0)$);
  \coordinate (C21) at ($(C7) + (0.3*\ml,0)$);
  \coordinate (C31) at ($0.5*(C20) + 0.5*(C21)$);
  \thrlines{C31};
  \ellipse{C21}{C20}{#5}{#6};
  
  \drawphotonline
}

\newcommand{\WFAcrPhMod}[7]{
  \coordinate (C1) at (-0.5*\ml,\maxh);
  \coordinate (C2) at (-0.5*\ml,\daspar*\maxh);
  \coordinate (C3) at (0.5*\ml,\daspar*\maxh);
  \coordinate (C4) at (-0.5*\ml,0);

  \coordinate (C5) at (0.5*\ml,0);
  \coordinate (C6) at (\ml,0);
  \pqlabel{C1}{C2}{C4};
  \coordinate (C10) at (\ml,\daspar*\maxh);
  \coordinate (C12) at (\ml,\maxh);
  \coordinate (C13) at (2.2*\ml,\maxh);
  \coordinate (C14) at (2.2*\ml,\daspar*\maxh);

  \draw[thick,#1] (C12) -- (C13);
  \draw[thick,#3] (C10) -- (C14);
  \coordinate (CM) at ($0.5*(C12)+0.5*(C14)-(0.2*\ml,0)$);
  \draw[fill=#4,draw=#7] (CM) circle ({0.5*(1-\daspar*\maxh)});

  \coordinate (C7) at (2.2*\ml,0);
  \draw[black,thick,#1] (C1) -- (C12);
  \draw[black,thick,#2] (C2) -- (C3) -- (C6) -- (C7);
  \draw[black,thick,#3] (C4) -- (C5) -- (C10);
  \coordinate (C20) at ($(C13) + (-0.1*\ml,0)$);
  \coordinate (C21) at ($(C7) + (0.3*\ml,0)$);
  \coordinate (C31) at ($0.5*(C20) + 0.5*(C21)$);
  \thrlines{C31};
  \ellipse{C21}{C20}{#5}{#6};

  \drawphotonline
}

\newcommand{\WFAPh}[7]{
  \coordinate (C1) at (-0.5*\ml, \maxh);
  \coordinate (C2) at (-0.5*\ml, \daspar*\maxh);
  \coordinate (C3) at (0.5*\ml,  \daspar*\maxh);
  \coordinate (C4) at (-0.5*\ml, 0);
  \pqlabel{C1}{C2}{C4};
  \coordinate (C10) at (\ml,\maxh-\mh);
  \coordinate (C12) at (\ml,\maxh);
  \coordinate (C13) at (2*\ml,\maxh);
  \coordinate (C14) at (2*\ml,\maxh-\mh);
  \lbox{C12}{C14}{#4}{#7};
  \coordinate (C7) at (2*\ml,0);
  \draw[black,thick,#1] (C1) -- (C12);
  \draw[black,thick,#2] (C2) -- (C3) -- (C10);
  \draw[black,thick,#3] (C4) -- (C7);
  \coordinate (C20) at ($(C13) - (0.2*\ml,0)$);
  \coordinate (C21) at ($(C7) + (0.2*\ml,0)$);
  \coordinate (C31) at ($0.5*(C20) + 0.5*(C21)$);
  \thrlines{C31};
  \ellipse{C21}{C20}{#5}{#6};
  
  \drawphotonline
}

\newcommand{\WFAcrPh}[7]{
  \coordinate (C1) at (-0.5*\ml,\maxh);
  \coordinate (C2) at (-0.5*\ml,\daspar*\maxh);
  \coordinate (C3) at (0.5*\ml,\daspar*\maxh);
  \coordinate (C4) at (-0.5*\ml,0);

  \coordinate (C5) at (0.5*\ml,0);
  \coordinate (C6) at (\ml,0);
  \pqlabel{C1}{C2}{C4};
  \coordinate (C10) at (\ml,\maxh-\mh);
  \coordinate (C12) at (\ml,\maxh);
  \coordinate (C13) at (2*\ml,\maxh);
  \coordinate (C14) at (2*\ml,\maxh-\mh);
  \lbox{C12}{C14}{#4}{#7};
  \coordinate (C7) at (2*\ml,0);
  \draw[black,thick,#1] (C1) -- (C12);
  \draw[black,thick,#2] (C2) -- (C3) -- (C6) -- (C7);
  \draw[black,thick,#3] (C4) -- (C5) -- (C10);
  \coordinate (C20) at ($(C13) - (0.2*\ml,0)$);
  \coordinate (C21) at ($(C7) + (0.2*\ml,0)$);
  \coordinate (C31) at ($0.5*(C20) + 0.5*(C21)$);
  \thrlines{C31};
  \ellipse{C21}{C20}{#5}{#6};

  \drawphotonline
}

\newcommand{\WFPhNL}[7]{
  \coordinate (C1) at (\ml,\maxh);
  \coordinate (C2) at (\ml,\daspar*\maxh);
  \coordinate (C4) at (\ml,0);

  \coordinate (C5) at (0.5*\ml,0);
  \coordinate (C6) at (\ml,0);
  \coordinate (C10) at (2*\ml,\daspar*\maxh);

  \coordinate (C13) at (2*\ml,\maxh);
  \coordinate (C7) at (2*\ml,0);
  \draw[black,thick,#1] (C1) -- (C13);
  \draw[black,thick,#2] (C2) -- (C10);
  \draw[black,thick,#3] (C4) -- (C7);
  \coordinate (C20) at ($(C13) - (0.25*\ml,0)$);
  \coordinate (C21) at ($(C7) + (0.25*\ml,0)$);
  \coordinate (C31) at ($0.5*(C20) + 0.5*(C21)$);
  \thrlines{C31};
  \coordinate (C32) at ($(C31 |- 0,0) + (0,-0.5*\maxh)$);
  \draw[black,thick,decorate,decoration=snake] (C31) -- (C32);
  \ellipse{C21}{C20}{#5}{#6};
}

\newcommand{\WFPh}[7]{
  \coordinate (C1) at (\ml,\maxh);
  \coordinate (C2) at (\ml,\daspar*\maxh);
  \coordinate (C4) at (\ml,0);

  \pqlabel{C1}{C2}{C4};
  \WFPhNL{#1}{#2}{#3}{#4}{#5}{#6}{#7}
}

\newcommand{\WFPhTnn}[8]{
  \coordinate (C01) at (0.25*\ml,\maxh);
  \coordinate (C02) at (0.25*\ml,\daspar*\maxh);
  \coordinate (C04) at (0.25*\ml,0);

  \coordinate (C1) at (1.25*\ml,\maxh);
  \coordinate (C2) at (1.25*\ml,\daspar*\maxh);
  \coordinate (C4) at (1.25*\ml,0);

  \pqlabel{C01}{C02}{C04};

  \draw[black,thick,#1] (C01) -- (C1);
  \draw[black,thick,#2] (C02) -- (C2);
  \draw[black,thick,#3] (C04) -- (C4);

  \coordinate (CM) at ($0.5*(C01)+0.5*(C2)$);
  \draw[fill=#8,draw=#8] (CM) circle ({0.5*(1-\daspar*\maxh)});

  \WFPhNL{#1}{#2}{#3}{#4}{#5}{#6}{#7}
}

\newcommand{\WFPhTnc}[8]{

  \coordinate (D1) at (0.5*\ml,0);
  \coordinate (D2) at (1.0*\ml,0);

  \coordinate (C41) at (1.0*\ml,\maxh);
  \coordinate (C42) at (1.0*\ml,\daspar*\maxh);
  \coordinate (C44) at (1.0*\ml,0);

  \coordinate (C31) at ($(C41)-(D1)$);
  \coordinate (C32) at ($(C42)-(D1)$);
  \coordinate (C34) at ($(C44)-(D1)$);

  \coordinate (C21) at ($(C31)-(D2)$);
  \coordinate (C22) at ($(C32)-(D2)$);
  \coordinate (C24) at ($(C34)-(D2)$);

  \coordinate (C11) at ($(C21)-(D1)$);
  \coordinate (C12) at ($(C22)-(D1)$);
  \coordinate (C14) at ($(C24)-(D1)$);

  \coordinate (C01) at ($(C11)-(D1)$);
  \coordinate (C02) at ($(C12)-(D1)$);
  \coordinate (C04) at ($(C14)-(D1)$);

  \pqlabel{C01}{C02}{C04};

  \draw[black,thick,#1] (C01) -- (C11) -- (C21) -- (C31) -- (C41);
  \draw[black,thick,#2] (C02) -- (C12) -- (C24) -- (C34) -- (C42);
  \draw[black,thick,#3] (C04) -- (C14) -- (C22) -- (C32) -- (C44);

  \coordinate (CM) at ($0.5*(C21)+0.5*(C32)$);
  \draw[fill=#8,draw=#8] (CM) circle ({0.5*(1-\daspar*\maxh)});

  \WFPhNL{#1}{#2}{#3}{#4}{#5}{#6}{#7}
}

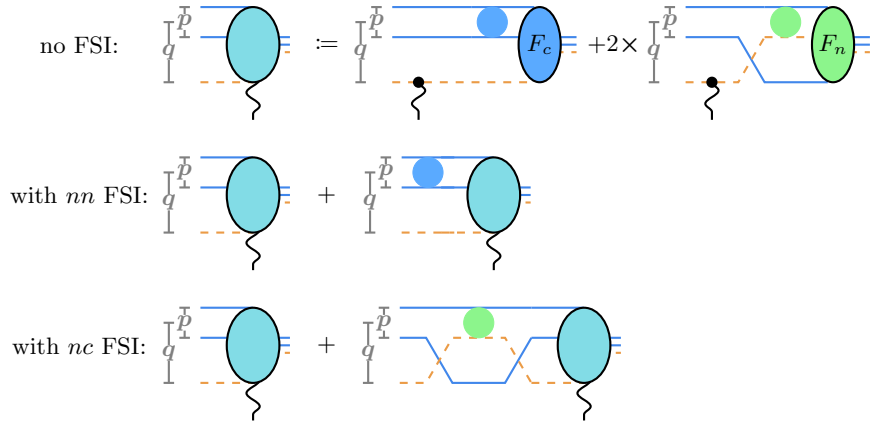
\begin{figure}[htb]
  \centering
  \begin{tikzpicture}[scale=1.0]
    \pgfmathsetmacro{\ml}{0.7};
    \pgfmathsetmacro{\mh}{0.1};
    \pgfmathsetmacro{\maxh}{10*\mh}

    \begin{scope}[yshift=2cm]
      \node at (-1.3*\ml, 0.5*\maxh) {no FSI:};
       
      \WFPh{\dnc}{\dnc}{dashed,\dcc}{\dnnc}{aqua1}{\(\)}{\dnnc}
      \node at (3.4*\ml,0.5*\maxh) {\(\coloneqq\)};

      \begin{scope}[xshift=3.6cm]
        \WFAPhMod{\dnc}{\dnc}{dashed,\dcc}{\dnnc}{\dacc}{\(F_c\)}{\dnnc}
        \node at (2.6,0.5*\maxh) {\(+2\cross\)};
        \begin{scope}[xshift=3.9cm]
          \WFAcrPhMod{\dnc}{\dnc}{dashed,\dcc}{\dncc}{\danc}{\(F_n\)}{\dncc}
        \end{scope}
      \end{scope}

    \end{scope}

    \begin{scope}[yshift=0cm]
      \node at (-1.3*\ml, 0.5*\maxh) {with \(nn\) FSI:};
      \WFPh{\dnc}{\dnc}{dashed,\dcc}{\dnnc}{aqua1}{\(\)}{\dnnc}
      \node at (3.4*\ml, 0.5*\maxh) {+};
      \begin{scope}[xshift=3.2cm]
        \WFPhTnn{\dnc}{\dnc}{dashed,\dcc}{\dnnc}{aqua1}{\(\)}{\dnnc}{\dnnc}
      \end{scope}
    \end{scope}

    \begin{scope}[yshift=-2cm]
      \node at (-1.3*\ml, 0.5*\maxh) {with \(nc\) FSI:};
      \WFPh{\dnc}{\dnc}{dashed,\dcc}{\dnnc}{aqua1}{\(\)}{\dnnc}
      \node at (3.4*\ml, 0.5*\maxh) {+};
      \begin{scope}[xshift=4.4cm]
        \WFPhTnc{\dnc}{\dnc}{dashed,\dcc}{\dnnc}{aqua1}{\(\)}{\dnnc}{\dncc}
      \end{scope}
    \end{scope}

  \end{tikzpicture}
  \caption{Diagrammatic representation of the \(E1\) matrix elements of distributions differing in the included FSIs.
  The neutrons are represented by blue solid lines and the \linine~core is represented by an orange dashed line.
  The first row describes the matrix element without FSI, whereby the ellipse with the external line on the left side
  represents the complete matrix element resulting from the action of the \(E1\) operator on the ground state.
  On the right-hand side of the first row this is made more explicit: The \(E1\) photons are represented by wiggly lines
  and the ground state is composed from its Faddeev amplitudes represented by ellipses with corresponding labels.
  The $nn$ and $nc$ t-matrices are represented by circles.
  The second row shows the contributions for the matrix element that includes \(nn\) FSI, while the third row describes the matrix element with \(nc\) FSI.}
  \label{fig:e1_diagrams}
\end{figure}

While they modify the shape of the \(E1\) distribution, the integral over the distribution, i.e.,
the overall \(E1\) strength, is conserved according to a sum rule and therefore is not affected by FSIs. First we 
explain how this sum rule comes about.

\subsection{The non-energy-weighted sum rule}

The cumulative $E1$ strength $B{(E1)}{(E)}$ is defined as the integral of the $E1$ strength up to an energy $E$. 
\begin{equation}
  B{(E1)}{(E)} \coloneqq \int_0^E \dd{E'} \frac{\dd{B(E1)}}{\dd{E'}}.
\end{equation}
According to the non-energy weighted sum rule (see, e.g., Ref. \cite{Forssen:2001rj}) the total strength, i.e., integrated all the way to infinite energy, is related to the RMS-radius \(\sqrt{\expval{r_c^2}}\)
by
\begin{equation}
  \lim_{E \to \infty} B{(E1)}{(E)} = \frac{3}{4\pi} Z_c^2 e^2 \expval{r_c^2} \,.
\end{equation}
This sum rule is derived using only the identity $r_c^2=\vec{r}_c \cdot \vec{r}_c$ and the completeness of the intermediate states. 
Therefore any approximate treatment of FSI should produce a cumulative distribution that has the same asymptotic value as that obtained when FSI is neglected, and that value should also be consistent with the $\expval{r_c^2}$ computed using the bound-state wave function. 

\subsection{\(E1\) strength distribution without FSI}
Our explicit expression for obtaining the \(E1\) strength of the \(2n\) halo nucleus with the ground state \(\ket{\Psi}\) reads
\begin{align}
  \frac{\dd B(E1)}{\dd E} &= e^2 Z_c^2 \sum_{\mu,M} \drint{\p}{\q} 
  \left| \imel{c}{ p,q; \zeta_c^{\K{1,\mu}}, \xi_c }{ r_c Y_{1\mu}{\K{\v{r_c}}} P_{\Xi_c}}{ \Psi }{} \right|^2 \de{E - \frac{p^2}{2\mu_{nn}} - \frac{q^2}{2\mu_{c(nn)}} }\,,
\end{align}
where we applied the approximation of using only the \(\ket{\Xi_c}_c\) partial-wave component.
This is realized by inserting the corresponding projection operator \(P_{\Xi_c}\).
It results in the omission of the \(l\neq 0\) components (see \cref{eq:Psi_c_l}).
This should be a good approximation since, as discussed above, the higher-\(l\) components are suppressed 
by a factor of at least 100.
The orbital angular momentum quantum numbers after the breakup are collected in the multiindex \(\zeta_c^{(1,\mu)}\),
which is given by
\begin{equation}
  \zeta_c^{(1,\mu)} =  \K{0,1} 1,\mu \,.
\end{equation}
Working in the $c$-representation for the wave function $|\Psi \rangle$ and retaining only the dominant component $\ibraket{c}{p,q;\zeta_c, \xi_c^{(M)}}{\Psi}{}$ it is straightforward to evaluate the operator $r_c Y_{1\mu}{\K{\v{r_c}}}$ in the plane-wave basis. This produces 
the concrete relation that is implemented:
\begin{align}
  \frac{\dd B(E1)}{\dd E} &=
    \frac{3 e^2 Z_c^2}{4\pi} \K{\frac{2}{A+2} }^2 
    \int_0^{\sqrt{2\mu_{c}E}} \dd{q} q^2 \sqrt{2\mu_{nn}^3} \sqrt{E-\frac{q^2}{2\mu_{c}}}  
    \left| \partial_{q'} \Psi_c{\K{\sqrt{2\mu_{nn}\K{E-\frac{q^2}{2\mu_{c}}}}, q'}} \Big|_{q'=q} \right|^2 \,,
    \label{eq:explicit_no_FSI}
\end{align}
where \(A\) is the mass number of the core and \(\mu_c \coloneqq \mu_{c(nn)}\) holds.
The wave function \(\Psi_c{\K{p,q}}\) is obtained from the Faddeev amplitudes as described in the previous section.

\subsection{\texorpdfstring{Including \(\boldsymbol{nn}\) FSI}{Includig nn FSI}}
\label{ssec:nn_fsi}

The dipole strength can also be calculated straightforwardly with \(nn\) final state interactions (FSI) taken into account.
This is done by inserting the M{\o}ller \cite{taylor_st,Moller:1945} operator of the \(nn\) interaction \(\Omega_{nn}^\dagger\) right before the final state:
\begin{equation}
  \frac{\dd B(E1)}{\dd E} = e^2 Z_c^2 \frac{1}{4} \sum_{\mu,M} \drint{\p}{\q} 
  \left| \imel{c}{ p,q; \zeta_c^{\K{1,\mu}},\xi_c^{(M)} }{ \Omega_{nn}^\dagger r_c Y_{1\mu}{\K{\v{r_c}}} P_{\Xi_c}}{ \Psi }{} \right|^2 \de{E - \frac{p^2}{2\mu_{nn}} - \frac{q^2}{2\mu_{c}} } \,.
\end{equation}
The M{\o}ller operator \(\Omega_{nn}^\dagger\) is given by
\begin{equation}
  \Omega_{nn}^\dagger =  \id + \drint{\p}{\q} \K{ \iketbra{p,q}{c}{p,q} \otimes \orbid \otimes \spid} t_{nn}{\K{E_p}} G_0^{(nn)}{\K{E_p}} \,.
\end{equation}
It converts the free state \(\ibra{c}{p,q;\zeta_c^{(1,\mu)},\xi_c}\) into the product of an $nn$ distorted wave and a plane-wave associated with the Jaocbi momentum of the core relative to the $nn$ pair
at \(t = 0\). The resulting three-body state has \(\ibra{c}{p,q;\zeta_c^{(1,\mu)},\xi_c} e^{\ci H_0 t}\) as an asymptotic state for
\(t \to \infty\).\footnote{The M{\o}ller operator thus makes use of the asymptotic condition, which requires that every state in the Hilbert space \(\mathcal{H}\) of solutions of a Schrödinger equation can form the asymptote of some scattering state, see e.g. Ref. \cite{taylor_st}.} The inclusion of $\Omega_{nn}^\dagger$ therefore ensures that the state obtained after the action of the $E1$ operator is overlapped with the three-body scattering state that includes $nn$ FSI, so leading to an \(E1\) distribution in which the effects of nn FSI are included. 

Since \(\Omega_{nn}^\dagger\) is an identity in the momentum of the spectator, \(q\), and in the associated parts of the partial wave states,
it commutes with the \(E1\) operator \(r_c Y_{1\mu}{\K{\v{r_c}}}\). This means that this calculation is an easy extension of the one described in the previous subsection. One obtains
\begin{align}
  \frac{\dd B(E1)}{\dd E} 
  &= \frac{3 e^2 Z_c^2}{4\pi} \K{\frac{2}{A+2} }^2 
  \int_0^{\sqrt{2\mu_{c}E}} \dd{q} q^2 \sqrt{2\mu_{nn}^3} \sqrt{E-\frac{q^2}{2\mu_{c}}}  
  \left| \partial_{q'} \Psi_c^{(\mathrm{wFSI})}{\K{\sqrt{2\mu_{nn}\K{E-\frac{q^2}{2\mu_{c}}}}, q'}} \Big|_{q'=q} \right|^2 \,.
\end{align}
where we have now defined a wave function that includes $nn$ FSI:
\begin{equation}
    \Psi_c^{(\mathrm{wFSI})}{\K{p, q}} \coloneqq \imel{c}{p, q; \zeta_c, \xi_c^{(M)} }{ \K{ \id + t_{nn}{\K{E_p}} G_0^{(nn)}{\K{E_p}} } }{\Psi}{} \,.
\end{equation}
An explicit expression is given in Eqs.~(35, 36) of Ref. \cite{Gobel:2021pvw}.
While the nucleus considered in Ref. \cite{Gobel:2021pvw} is \({}^6\)He,
the corresponding equations apply here as well, as they only describe
the inclusion of \(nn\) FSI.

\subsection{\texorpdfstring{Including \(\boldsymbol{nc}\) FSI}{Including nc FSI}}

The distribution with \(nc\) FSI can be obtained in a similar fashion as the distribution with \(nn\) FSI.
The M{\o}ller operator \(\Omega_{nn}^\dagger\) has to be replaced by \(\Omega_{nc}^\dagger\), which is given by
\begin{equation}
  \label{eq:ncFSI}
  \Omega_{nc}^\dagger =  \id + \drint{\p}{\q} \K{ \iketbra{p,q}{n}{p,q} \otimes \orbid \otimes \spid} t_{nc}{\K{E_p}} G_0^{(nc)}{\K{E_p}} \,.
\end{equation}
However, because \(\Omega_{nc}^\dagger\) does not commute with the \(E1\)
operator, multiple three-body bases have to be used in the evaluation of
Eq.~(\ref{eq:ncFSI}). 
An explicit expression for the distribution with \(nc\) FSI will be
given below.

\subsection{Results}

We now show our results for the \(E1\) strength distributions from calculations with the two \(nc\) interaction channels (\(s=\)1 and \(s\)=2).
For the spatial part of the solution, we employed the equivalence statement described in \cref{ap:equiv_one_channel_two_channels}, i.e., we did an calculation with core spin and overall spin set to zero.

\begin{figure}[htb]
  \centering
  \includegraphics[width=0.45\textwidth]{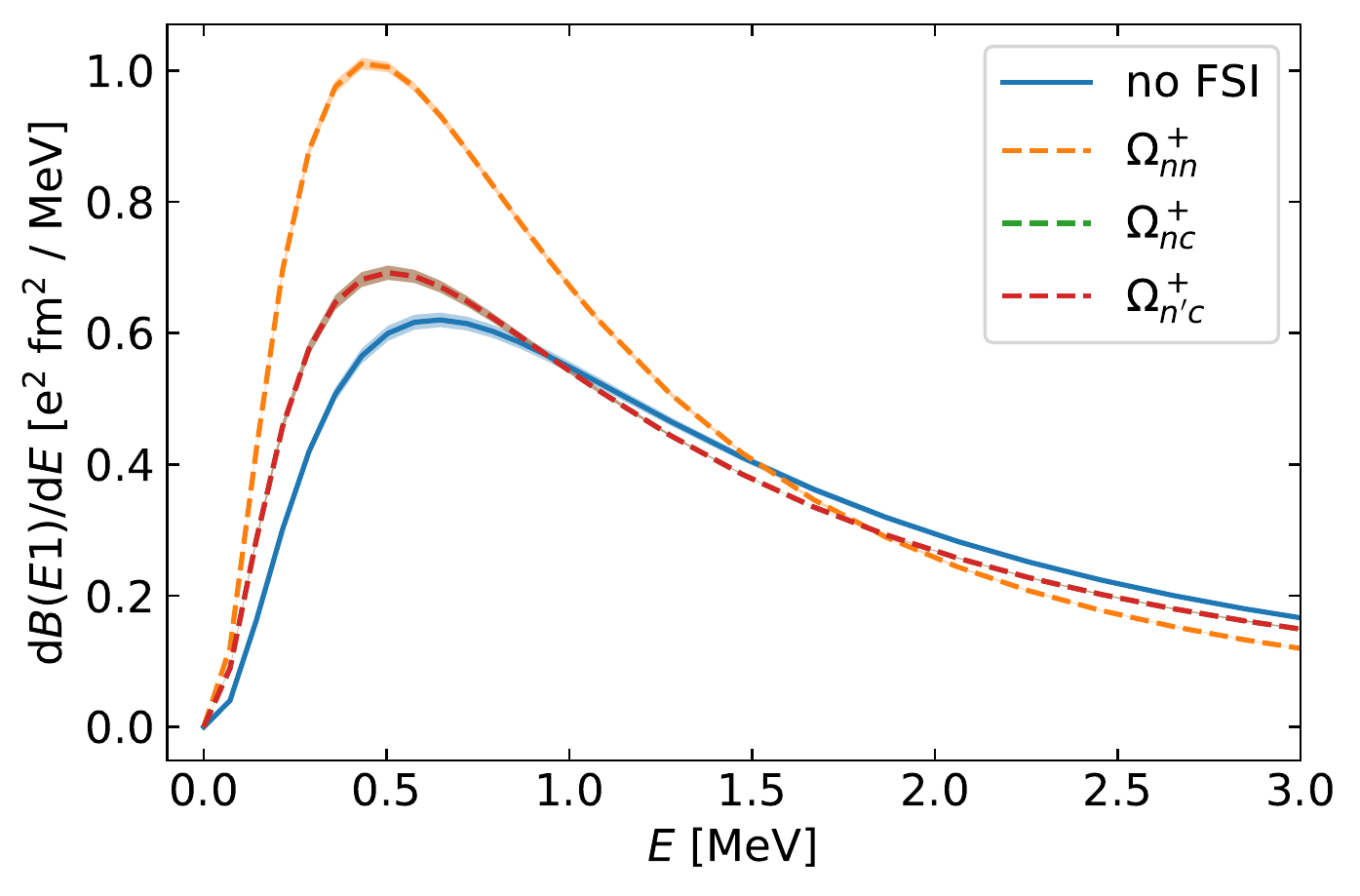}
  \includegraphics[width=0.45\textwidth]{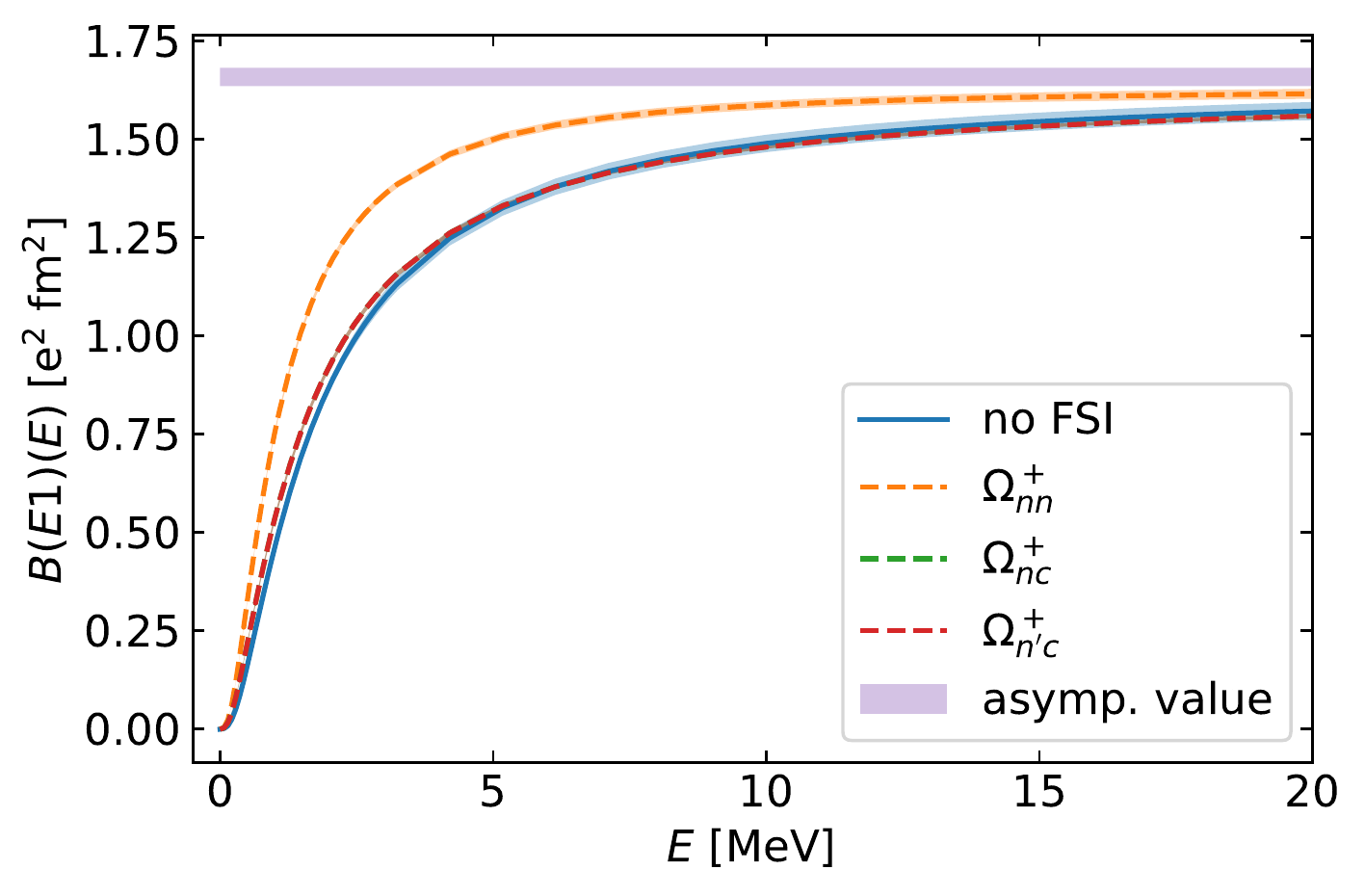}
  \caption{The left panels shows \(E1\) strength distributions of \lieleven~with different FSIs included.
  The right panel shows the corresponding cumulative \(E1\) strength distributions.
  Numerical uncertainties are indicated by bands, which are very narrow here.
  They were obtained by comparing the calculations with ones having roughly two thirds as many mesh points
  and a cutoff of three fourths of the original one.}
  \label{fig:e1_distribs_sg_fsi}
\end{figure}

The results for the \(E1\) distributions with no FSI as well with either a single \(nn\) or \(nc\) FSI are shown
in the left-hand panel of \cref{fig:e1_distribs_sg_fsi}. In the right panel, we show
the cumulative distribution $B(E1)(E)$. 

It can be seen that the \(nn\) FSI influences the shape of the strength distribution significantly, producing a strong enhancement at low energies, and a continuing depletion of the strength at higher energies.  \(nc\) FSI (and \(n'c\) FSI) also increase the strength at low energy, but their influence is markedly less than that of the 
 \(nn\) FSI.

In the case of the cumulative distributions a common asymptotic value can be observed, in accordance with the sum rule. 
We expect that the agreement would  become even better if we continued the calculation to higher energies.
The asymptotic values are also in approximate agreement with the overall \(E1\) strength value calculated from
\(\expval{r_c^2}\) computed in Sec.~\ref{sec:radius}

\subsection{\texorpdfstring{Role of \(\boldsymbol{nc}\) interaction channels in \(\boldsymbol{{}^{11}}\)Li}{Role of nc interaction channels in Li-11}}

\begin{figure}[htb]
  \centering
  \includegraphics[width=0.9\textwidth]{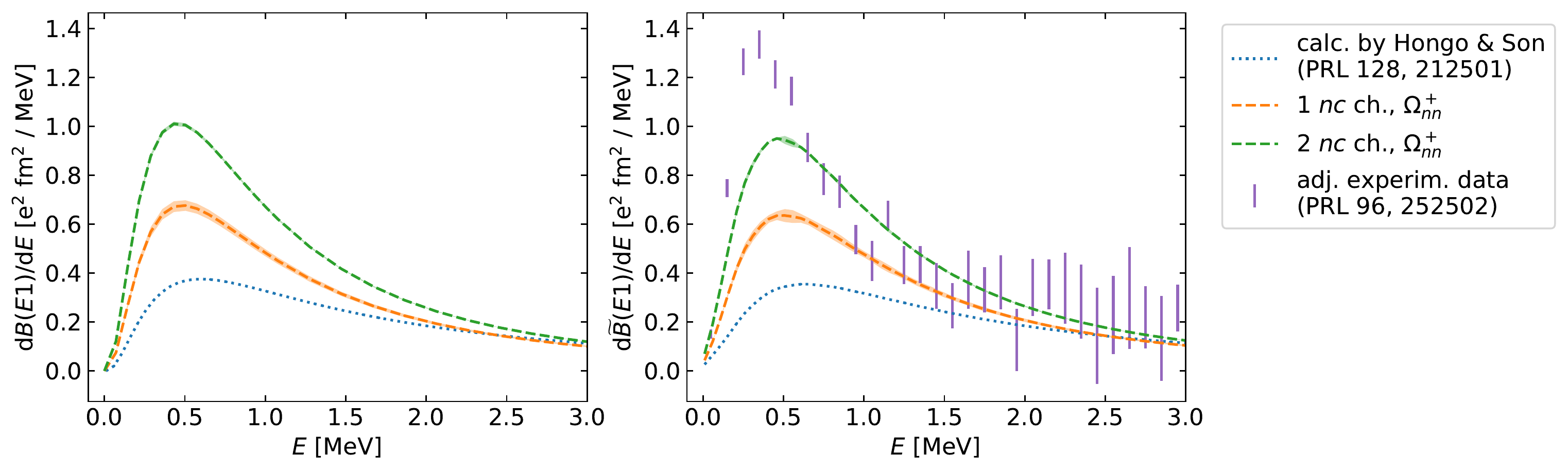}
  \caption{\(E1\) strength distributions with \(nn\) FSI included and different numbers of \(nc\) interaction channels for
  the ground state. We show the result by Hongo and Son~\cite{Hongo:2022sdr} (blue), which corresponds to no \(nc\) interaction spin channels, in comparison with our results using one spin channel (orange) and two spin channels (green).
  The left panel shows the theoretical curves. In the right panel these distributions have been folder with the detector resolution and compared to the experimental data from Nakamura et al. \cite{Nakamura:2006zz} (adjusted to the current \(S_{2n}\) value).}\label{fig:e1_distribs_nc_ch_cmp}
\end{figure}

Now that we have assessed the impact of FSI on the results, 
we want to compare the calculation with two \(nc\) interaction channels  
to that with only one \(nc\) interaction channel. 
The parameter describing the \(nc\) interaction \(E_{nc}\) was in both cases the same, the difference is that when only the $s=2$ \(nc\) spin channel is active the interaction is switched off in the $s=1$ \(nc\) spin channel.  
The results of the calculations are shown in \cref{fig:e1_distribs_nc_ch_cmp}.
While the left panel contains the theoretical curves, the right panels contains the same distributions but folded with the detector resolution
and compared with the experimental data.
More information on the folding and the experimental data can be found in \cref{ssec:cmp_expm}.

Also shown there is a recent calculation of Hongo and Son for \(2n\)
halo nuclei~\cite{Hongo:2022sdr}.
In this context, it is important to note that the universal curve from Hongo and Son is derived in an EFT picture of \(2n\)-halo nuclei in which they are bound by the \(nn\) interaction
and a three-body force: the \(nc\) interaction is taken to be an NLO effect there. 
It therefore applies to \(2n\) halo nuclei where \(S_{2n}\) and \(\epsilon_n = \hbar^2 /(2\mu_{nn} a_{nn}^2)\) are smaller than
all other energy scales, and in particular smaller than \(\epsilon_{nc}\). This is not the case in \lieleven~due to the near-threshold resonance in \({}^{10}\)Li, and 
Hongo and Son themselves say the applicability of their results to \lieleven~is ``doubtful''. Fig.~\ref{fig:e1_distribs_nc_ch_cmp}
shows that the Hongo and Son calculation predicts a much lower $E1$ strength than any of the calculations in which a low-energy 
\(nc\) virtual state plays a role in the structure of \({}^{11}\)Li.

Since \(nn\) FSI is included in all curves, they can directly be compared.
It is clear that the low-energy strength increases with the number of \(nc\) interaction channels.
The result by Hongo and Son~\cite{Hongo:2022sdr} (blue curve) has too little strength for $E \approx 0.5$ MeV. The calculation with one channel (orange curve) has already more strength,
while using two channels (green curve) results in the highest strength.
Since the \(nc\) interaction does not appear in any of the final-state-interaction treatments used here all the differences between the different results stem from effects in the initial-state \(^{11}\)Li nucleus. Crucially, all three calculations are adjusted to the same \(S_{2n}\). 
It is then quite striking that the $E1$ strength increases appreciably depending on the fraction of \(nc\) pairs that interact with a large \(nc\) scattering length: 0 of them, 5/8 of them, or all of them.
(The factor of 5/8 is the ratio of the spin multiplicity of the \(nc\) interaction channel and the sum of multiplicities of all
possible \(nc\) spin couplings, see \cref{eq:recoupling_pw_gs}.)

The description using two \(nc\) interaction channels, in which all the \(nc\) pairs in \({}^{11}\)Li can scatter via a large \(a_{nc}\),
yields a much better description of the data in this leading-order calculation. 
Therefore we will use it for the investigations of the next section, where we seek to include effects due to both \(nn\) and \(nc\) FSI.

\section{Sum-rule preserving approximation schemes for FSI}
\label{sec:higherorder}

In this section, we explore different approximation schemes for the
FSI in detail. Our goals are:
\begin{enumerate}
  \item[(i)] derive accurate approximation schemes
    for practical calculations, and
  \item[(ii)] understand the role of different
    FSI channels and orders in the multiple scattering series.
\end{enumerate}
For this purpose, we make use of M{\o}ller operators, which were already briefly discussed in
  \cref{ssec:nn_fsi}.
  The full final state can be written as
  \begin{equation}
    \ibra{c}{p,q;\Xi_f}{\Omega^\dagger_{nn+nc+n'c}} \,,
  \end{equation}
  where \(\Xi_f\) is some set of orbital angular momentum and spin quantum numbers and \(\Omega_{nn+nc+n'c}^\dagger\)
  is the M{\o}ller operator containing all the two-body final-state interactions:
  \begin{equation}
    \Omega_{nn + nc + n'c} = \id + \drint{\p}{\q} \sum_{\Xi} \frac{1}{E_{p,q} - H_0 - V_{nn} - V_{nc} - V_{n'c} + \ci \epsilon} \K{V_{nn} + V_{nc} + V_{n'c}} \iketbra{p,q;\Xi}{c}{}.
  \end{equation}
  Calculating the action of this operator on a plane-wave state is challenging due to the presence of three different two-body potentials,
  $V_{nn}, V_{nc}, V_{n'c}$. 
  To obtain the three-body scattering state, we would have to solve the Faddeev (or equivalent) equations above three-body breakup. 
  Therefore we are interested in approximation strategies, especially since comparing different approximations can lead to additional
  insights into the final-state dynamics.
  Note, however, that there are also calculations based on full three-body scattering states of \lieleven~available, see, e.g., Ref. \cite{Kikuchi:2013ula}.

  We continue by analyzing the final scattering state in order to obtain approximations.
  Using the Faddeev equations for scattering states as an intermediate step produces
  \begin{equation}\label{eq:mss}
    \ibra{c}{p,q; \Xi_f} \Omega_{nn + nc + n'c}^\dagger = \ibra{c}{p,q; \Xi_f} \K{ \id + \sum_i t_i G_0 + \sum_i t_i G_0 \sum_{j \neq i} t_j G_0 + ...} \,,
  \end{equation}
  where we omitted the arguments of the t-matrices and Green's functions for brevity.
  From this the following approximation can be obtained:
  \begin{equation}
    \ibra{c}{p,q; \Xi_f} \K{ \id + \sum_i t_i G_0 } \,.
  \end{equation}
  This treatment, which keeps the first-order terms in the multiple-scattering series, is not unitary. 
  In contrast the M{\o}ller operators introduced in the previous section are isometric and unitary (since neither the \(nn\) nor the \(nc\) subsystem supports a bound state).
  Non-unitarity can lead to unphysical gains and losses of probability, which are manifest as violations of the non-energy-weighted sum rule.

  In order to ensure we have a sum-rule-preserving approximation scheme, 
we propose to use products of M{\o}ller operators, whereby the single M{\o}ller operators correspond to single types
  of interactions (\(nn\) or \(nc\) or \(n'c\)).
This ensures that we keep unitarity-preserving combinations of terms in the multiple-scattering series. Of course, in doing so 
  we do not truncate the multiple-scattering series at a given order in t-matrices, because it is not possible to do that and also maintain unitarity. Unitarity is only obtained in such a scheme if the multiple-scattering series is summed to infinite order. 

\subsection{Organization of FSI calculations}
\label{ssec:calc_strategy}

We will now work out how to efficiently organize calculations of $E1$ distributions with FSIs
based on combinations of M{\o}ller operators.
We will identify ingredients which different distributions have in common
and describe the calculation of the different matrix elements on this basis. 
The procedure to obtain the final distributions from the matrix elements is then basically independent of included FSIs.

In proceeding in this way, it is useful to specify the initial and final states and to discuss their partial-wave structure.
The initial state used in the calculations of this section is that obtained by acting with the $E1$ operator on the \(\Xi_c \coloneqq \zeta_c, \xi_c \) partial-wave component of the ground state:
\begin{equation}
  \ket{i} \coloneqq \mathcal{M}{\K{\mathrm{E1},\mu}} P_{\Xi_c} \ket{\Psi}.
\end{equation}
After this $E1$ transition (FSIs not yet included) the system is in the partial-wave state
\begin{equation}
  \iket{\zeta_c^{\K{1,\mu}},\xi_c}{c} \,.
\end{equation}

In order to compactly specify the final states after FSIs, which can be in various partial waves due to recoupling,
we introduce the multiindex
\begin{equation}
  \iket{ \zeta_f^{\K{\lb,\lbb;\mu}} }{c} \coloneqq \iket{ \K{\lb,\lbb} 1,\mu }{c} 
\end{equation}
for the spatial part and the multiindex
\begin{equation}
  \iket{ \xi_f^{\K{\lsb;M}} }{c} \coloneqq \iket{ \K{\lsb, \frac{3}{2}} \frac{3}{2},M }{c} 
\end{equation}
for the spin part.
For illustrative purposes, we put the multiindices directly into kets, 
since they are usually used with the core as spectator.
The quantum numbers here have bars on top in order to distinguish them from the ones characterizing the ground state.
While overall spin and orbital angular momentum are conserved, the subsystem quantum numbers are in general not conserved.

In the case of the two \(nc\) interaction channels (\(s_c+1/2\) and \(s_c-1/2\)), the Hamilton operator decouples into one with the
\(nn\) system in spin 0 configuration and one with the \(nn\) system in spin 1 configuration.
Therefore, the initial state with \(s = 0\) will remain in this configuration and we have \(\bar{s} = 0\).

On this basis we define the following ``ingredients'':
\begin{align}
  \maf{1}{\lb,\lbb;\mu;\lsb,M}{p,q} &\coloneqq \imel{c}{p,q; \zeta_{f}^{\K{\lb,\lbb;\mu}}, \xi_f^{\K{\lsb;M}} }{ \K{ \Omega_{nc}^\dagger - \id} }{i}{} \,, \label{eq:def_maf1} \\
  \maf{2}{\lb,\lbb;\mu;\lsb,M}{p,q} &\coloneqq \imel{c}{p,q; \zeta_{f}^{\K{\lb,\lbb;\mu}}, \xi_f^{\K{\lsb;M}} }{ \K{ \Omega_{n'c}^\dagger -\id} \K{ \Omega_{nc}^\dagger - \id} }{ \vphantom{\Omega_{f}^{\K{\lb,\lbb;\mu}}} i}{} \,. \label{eq:def_maf2}
\end{align}
To evaluate these matrix elements we must recouple the partial-wave states (momenta, angular momenta, and spins) from the $c$-spectator basis to the $n'$-spectator basis
in the case of $\maf{1}{\lb,\lbb;\mu;\lsb,M}{p,q}$, and then, additionally from the $n'$-spectator basis to the $n$-spectator basis
in the case of  $\maf{2}{\lb,\lbb;\mu;\lsb,M}{p,q}$. 
In both cases the final state is specified using the core as spectator implying another recoupling. 
The details of those calculations, together with explicit expressions for 
these matrix elements, are given in \cref{ap:mc_A_1} and \cref{ap:mc_A_2} respectively. 

Once $\maf{1}{\lb,\lbb;\mu;\lsb,M}{p,q}$ and $\maf{2}{\lb,\lbb;\mu;\lsb,M}{p,q}$ have been calculated the related matrix elements
\begin{align}
  \maft{1}{\lb,\lbb;\mu;\lsb,M}{p,q} &\coloneqq \imel{c}{p,q; \zeta_{f}^{\K{\lb,\lbb;\mu}}, \xi_f^{\K{\lsb;M}} }{ \K{ \Omega_{n'c}^\dagger - \id} }{i}{}  \,, \\
  \maft{2}{\lb,\lbb;\mu;\lsb,M}{p,q} &\coloneqq \imel{c}{p,q; \zeta_{f}^{\K{\lb,\lbb;\mu}}, \xi_f^{\K{\lsb;M}} }{ \K{ \Omega_{nc}^\dagger -\id} \K{ \Omega_{n'c}^\dagger - \id} }{ \vphantom{\Omega_{f}^{\K{\lb,\lbb;\mu}}} i}{} \,,
\end{align}
in which the roles of the two neutrons, $n$ and $n'$, have been interchanged, can be found  
using the properties of the permutation operators \(\pmospatial\) and \(\pmospin\). These
yield the following relations between the \(\mathcal{A}\) and \(\widetilde{\mathcal{A}}\) functions:
\begin{align}\label{eq:maft1_sym}
  \maft{1}{\lb,\lbb;\mu;\lsb,M}{p,q} &= \K{-1}^{\lb}  \K{-1}^{-\lsb}       \maf{1}{\lb,\lbb;\mu;\lsb,M}{p,q} \,, \\
  \maft{2}{\lb,\lbb;\mu;\lsb,M}{p,q} &= \K{-1}^{-\lb} \K{-1}^{-\K{1-\lsb}} \K{-1} \maf{2}{\lb,\lbb;\mu;\lsb,M}{p,q} \,,
\end{align}
i.e., the tilde matrix elements are the same as the unbarred ones up to phase factors stemming from \(nn\) permutations.

Another important ingredient is the overlap of final and initial state with no ``FSI operator" in between:
\begin{equation}
  \maf{0}{\mu}{p,q} \kd{\lb}{0} \kd{\lbb}{1} \coloneqq \ibraket{c}{p,q; \zeta_{f}^{\K{\lb,\lbb;\mu}}, \xi_c^{(M)}  }{i}{}  \,.
\end{equation}
This was already evaluated as part of the calculation of \cref{eq:explicit_no_FSI}. 
This function depends on \(\mu\) but not on \(\lb\), \(\lbb\) and \(\bar{s}\), since the overlap
on the right is non-vanishing only if \(\lb = 0\) and \(\lbb = 1\) and \(\bar{s} = 0\).

Using these ingredients, and the definition 
\begin{equation}
  \widetilde{\Omega}_{ij} \coloneqq \Omega_{ij} - \id \,,
\end{equation}
we can obtain comparatively compact expressions for the matrix elements of different
combinations of M{\o}ller operators.

First, we see that, with $  \maf{0}{\mu}{p,q}$ in hand, the matrix element of the \(nn\) M{\o}ller operator, implicitly worked out in the previous section, is easily written as\footnote{
  Note that \(\tau\) which is the "reduced t-matrix element" takes sometimes a momentum and sometimes an energy as argument in this paper.
  This variation stems from the context and there is no other reason. \(\tau_k{\K{E}}\) can be read as \(\tau_k{\K{\sqrt{2\mu_{ij}E}}}\).
}
:
\begin{align}
  & \imel{c}{ p,q; \zeta_{f}^{\K{\lb,\lbb;\mu}}, \xi_f^{\K{\lsb;M}} }{ \Omega_{nn}^\dagger  }{i}{} =
  \imel{c}{ p,q; \zeta_{f}^{\K{\lb,\lbb;\mu}}, \xi_f^{\K{\lsb;M}} }{ \K{ \id + \drint{\pt}{\qt} \sum_\Omega \iketbra{\pt,\qt;\Omega}{c}{...} t_{nn}{\K{E_{\pt}}} G_0^{(nn)}{\K{E_{\pt}}} }  }{i}{} \\
  &\quad = \kd{\lb}{0} \kd{\lbb}{1} \kd{\lsb}{0} \K{ \maf{0}{\mu}{p,q} + \frac{2}{\pi} g_{0}{\K{p}} \tau_{nn}{\K{p}} \rint{\pp} g_0{\K{\pp}} \K{p^2 -\pp[2] + \ci \epsilon}^{-1} \maf{0}{\mu}{\pp,q} } \\
  &\quad \eqqcolon \mbf{\mu}{p,q} \kd{\lb}{0} \kd{\lbb}{1} \kd{\lsb}{0} \,. \label{eq:one_mo_mel}
\end{align}

Then, since we also have a result for $  \maft{1}{\lb,\lbb;\mu;\lsb,M}{p,q} $, if we notate the action of the \(nn\) M{\o}ller operator to be the
 \(\mbf{\mu}{p,q}\) defined in \cref{eq:one_mo_mel}, we can write the matrix element of the product of the \(nn\) and \(nc\)
M{\o}ller operators as:
\begin{align}
  &\imel{c}{ p,q; \zeta_{f}^{\K{\lb,\lbb;\mu}}, \xi_f^{\K{\lsb;M}} }{ \Omega_{nn}^\dagger \Omega_{nc}^\dagger  }{i}{} =
  \imel{c}{ p,q; \zeta_{f}^{\K{\lb,\lbb;\mu}}, \xi_f^{\K{\lsb;M}} }{ \K{ \Omega_{nn}^\dagger + \widetilde{\Omega}_{nc}^\dagger + \widetilde{\Omega}_{nn}^\dagger \widetilde{\Omega}_{nc}^\dagger}  }{i}{} \\
  &\quad = \kd{\lb}{0}\kd{\lbb}{1} \kd{\lsb}{0} \mbf{\mu}{p,q} + \maf{1}{\lb,\lbb;\mu;\lsb,M}{p,q} \nonumber \\
  &\quad\quad + \kd{\lb}{0}\kd{\lbb}{1} \kd{\lsb}{0} \frac{2}{\pi} g_{0}{\K{p}} \tau_{nn}{\K{p}} \rint{\pp} g_0{\K{\pp}} \K{p^2 -\pp[2] + \ci \epsilon}^{-1} \maf{1}{0,1;\mu;0,M}{\pp,q} \,. \label{eq:two_mo_mel}
\end{align}

Finally, we can write the matrix element of a product of three M{\o}ller operators
\begin{align}\label{eq:three_mo_mel}
  & \imel{c}{ p,q; \zeta_{f}^{\K{\lb,\lbb;\mu}}, \xi_f^{\K{\lsb;M}} }{ \Omega_{nn}^\dagger \Omega_{n'c}^\dagger \Omega_{nc}^\dagger  }{i}{} =
    \kd{\lb}{0}\kd{\lbb}{1} \kd{\lsb}{0} \mbf{\mu}{p,q} + \K{1 + \K{-1}^{\lb + \lsb}} \maf{1}{\lb,\lbb;\mu;\lsb,M}{p,q} \nonumber \\
  &\quad + 2 \kd{\lb}{0}\kd{\lbb}{1} \kd{\lsb}{0} \frac{2}{\pi} g_{0}{\K{p}} \tau_{nn}{\K{p}} \rint{\pp} g_0{\K{\pp}} \K{p^2 -\pp[2] + \ci \epsilon}^{-1} \maf{1}{0,1;\mu;0,M}{\pp,q} \nonumber \\
  &\quad + \maf{2}{\lb,\lbb;\mu;\lsb,M}{p,q} + \kd{\lb}{0} \kd{\lbb}{1} \kd{\lsb}{0} \frac{2}{\pi} g_{0}{\K{p}} \tau_{nn}{\K{p}} \rint{\pp} g_0{\K{\pp}} \K{p^2 -\pp[2] + \ci \epsilon}^{-1} \maf{2}{0,1;\mu;0,M}{\pp,q} \,,
\end{align}
where we also used \cref{eq:maft1_sym}.
In the case of \cref{eq:three_mo_mel} the relation
\begin{align}
  \Omega_{nn}^\dagger \Omega_{n'c}^\dagger \Omega_{nc}^\dagger &= \K{\id + \widetilde{\Omega}_{nn}^\dagger} \K{\id + \widetilde{\Omega}_{n'c}^\dagger} \K{\id + \widetilde{\Omega}_{nc}^\dagger} \\
  &= \Omega_{nn}^\dagger + \K{ \widetilde{\Omega}_{n'c}^\dagger + \widetilde{\Omega}_{nc}^\dagger } + \widetilde{\Omega}_{nn}^\dagger \K{ \widetilde{\Omega}_{n'c}^\dagger + \widetilde{\Omega}_{nc}^\dagger }
  + \widetilde{\Omega}_{n'c}^\dagger \widetilde{\Omega}_{nc}^\dagger + \widetilde{\Omega}_{nn}^\dagger \widetilde{\Omega}_{n'c}^\dagger \widetilde{\Omega}_{nc}^\dagger 
\end{align}
was employed.
Note that, with a result for the product of three M{\o}ller operators in hand, the  expression for the matrix element of \(\Omega_{n'c}^\dagger \Omega_{nc}^\dagger\) can be obtained from
\cref{eq:three_mo_mel} by replacing \(\mathcal{B}_\mu{\K{p,q}}\) by \(\mathcal{A}_\mu^{(0)}{\K{p,q}}\) and setting
the \(\tau_{nn}\) in this formula to zero for all momenta.

In each of these expressions integrals stemming from taking the t-matrix elements have to evaluated.
Limiting ourselves to the case of sharp-cutoff regularization at \(\Lambda\) in the subsystems, i.e.,
\begin{equation}
  g_l{\K{p}} \coloneqq p^l \Theta{\K{\Lambda - p}} \,,
\end{equation}
we can use the relation
\begin{align}
  \rint{\pp} \frac{ g_0{\K{\pp}} f{\K{\pp,q}} }{p^2 -\pp[2] + \ci \epsilon} =
    \int_0^\Lambda \dd{\pp} \frac{ \pp[2] f{\K{\pp, q}} - p^2 f{\K{p, q}} }{p^2 -\pp[2]} - \K{\frac{\ci \pi}{2} - \frac{1}{2}\ln{\K{\frac{\Lambda + p}{\Lambda - p}}}} g_0{\K{p}} p f{\K{p,q}} \,.
\end{align}
For a derivation see the supplemental material of Ref. \cite{Gobel:2021pvw}.

\subsection{Physical properties of approximation schemes}

Now we are in a position to calculate the $E1$ strength distribution for all these additional combinations of M{\o}ller operators.
The combinations we are interested in are listed in \cref{tab:fsi_schemes}.
In the third, fourth, and fifth columns some desirable properties of the resulting matrix element are listed.
A particularly important one is the unitarity of the approach:  the physical FSIs are norm-preserving, because there is no ``probability flow'' into bound states in our problem. That means that any violation of unitarity represents a defect of our approximation scheme. 
Another relevant aspect is if the FSI operator commutes with the \(nn\) permutation operator \(\pmo\).
The full FSI operator commutes with \(\pmo\), so that the \(nn\) antisymmetry is not broken by FSI.
Also here, any violation of \(nn\) anti-symmetry must result from approximations we have introduced. 
Additional characteristics are the order of the expression in the t-matrices, as well as whether all different
two-body interactions up to that order are taken into account.
Note that in the table also some abbreviations for the different combinations of operators are introduced: a bar over the subscript $n$ means that the operator has been averaged between the two identical neutrons, so as to ensure \(nn\) antisymmetry. The bar over $\Omega_3^\dagger$ in $\bar{\Omega}_3^\dagger$ indicates that the two different orderings of the M{\o}ller operators $\Omega_{nc}^\dagger$ and $\Omega_{n'c}^\dagger$ have been averaged. And in both $(\Omega_3')^\dagger$ and $(\bar{\Omega}_3')^\dagger$ the prime indicates that the $nc$ and $n'c$ interactions come after the $nn$ interaction, rather than before it.

\newcommand{\ttco}[1]{\textcolor[HTML]{2B72E3}{#1}}
\newcommand{\ttct}[1]{\textcolor[HTML]{4DBD53}{#1}}

\newcolumntype{C}{>{\centering\arraybackslash}X}%
\begin{table}[H]
  \centering
  \caption{Overview of different FSI schemes specified in terms of the used combinations of M{\o}ller operators.
  Especially interesting combinations are highlighted in color.}
  \label{tab:fsi_schemes}
  \begin{tabularx}{0.7\textwidth}{cCCCC}
    \toprule
    operator & max. order\newline in \(t_{ij} G_0\) & commutes\newline with \(\pmo\) & unitary & all two-body\newline interactions included\\
    \midrule
    \(\id\) & 0 & \checkm & \checkm & \crossm \\
    \ttco{\(\K{\Omega^{(\mathrm{fo})}}^\dagger\)} & \ttco{1} & \ttco{\checkm} & \ttco{\crossm} & \ttco{\checkm} \\
    \(\Omega_{nn}^\dagger\)                & 1 & \checkm & \checkm & \crossm \\
    \(\Omega_{nc}^\dagger\)                & 1 & \crossm & \checkm & \crossm \\
    \(\Omega_{\bar{n}c}^\dagger \coloneqq \frac{1}{2}\K{ \Omega_{nc}^\dagger + \Omega_{n'c}^\dagger }\)          & 1 & \checkm & \crossm & \crossm \\
    \(\Omega_{nn}^\dagger \Omega_{nc}^\dagger\)       & 2 & \crossm & \checkm & \crossm \\
    \(\Omega_{nn}^\dagger \Omega_{\bar{n}c}^\dagger\) & 2 & \checkm & \crossm & \checkm \\
    \ttct{\(\Omega_3^\dagger \coloneqq \Omega_{nn}^\dagger \Omega_{n'c}^\dagger \Omega_{nc}^\dagger \)} & \ttct{3} & \ttct{\crossm} & \ttct{\checkm} & \ttct{\checkm} \\ 
    \ttct{\(\bar{\Omega}_3^\dagger \coloneqq \frac{1}{2} \Omega_{nn}^\dagger \K{ \Omega_{n'c}^\dagger \Omega_{nc}^\dagger + \Omega_{nc}^\dagger \Omega_{n'c}^\dagger} \)} & \ttct{3} & \ttct{\checkm} & \ttct{\crossm} & \ttct{\checkm} \\ 
    \ttct{\(\K{\Omega_3^\prime}^\dagger \coloneqq \Omega_{n'c}^\dagger \Omega_{nc}^\dagger \Omega_{nn}^\dagger \)} & \ttct{3} & \ttct{\crossm} & \ttct{\checkm} & \ttct{\checkm} \\ 
    \ttct{\(\K{\bar{\Omega}_3^\prime}^\dagger \coloneqq \frac{1}{2} \K{ \Omega_{n'c}^\dagger \Omega_{nc}^\dagger + \Omega_{nc}^\dagger \Omega_{n'c}^\dagger} \Omega_{nn}^\dagger \)} & \ttct{3} & \ttct{\checkm} & \ttct{\crossm} & \ttct{\checkm} \\ 
    \bottomrule
  \end{tabularx}
\end{table}

The table makes it clear that the different combinations have different advantages.
The expression using all t-matrices up to first order, i.e. the expression using \(\K{\Omega^{(\mathrm{fo})}}^\dagger\), has
the advantage that \(nn\) antisymmetry is preserved and all interactions are taken into account.
However, it is not necessarily unitary.
The combination of all three different M{\o}ller operators \(\Omega_3^\dagger\) has the advantages of taking all interactions into account
and of being unitary.
However, it does not commute with \(\pmo\).
It is possible to produce a commutative variant of this combination called \(\bar{\Omega}_3^\dagger\) at the price of losing guaranteed unitarity.
In terms of this selection of ``features'' it is thereby on a par with \(\K{\Omega^{(\mathrm{fo})}}^\dagger\).
However, on a quantitative level there might be significant differences: It might be that the violation of unitarity of \(\bar{\Omega}_3^\dagger\) is much smaller than in the case of \(\K{\Omega^{(\mathrm{fo})}}^\dagger\).

Before showing the results, we want to mention that in the calculation of the \(E1\) distributions from the matrix elements
sums over final-state quantum numbers are involved.
In the case of some terms truncations are necessary.
A detailed discussion of these sums and the convergence of the truncation can be found in \cref{ap:qnsum_trunc}.

\subsection{Numerical results}

\begin{figure}[htb]
  \centering
  \includegraphics[width=0.45\textwidth]{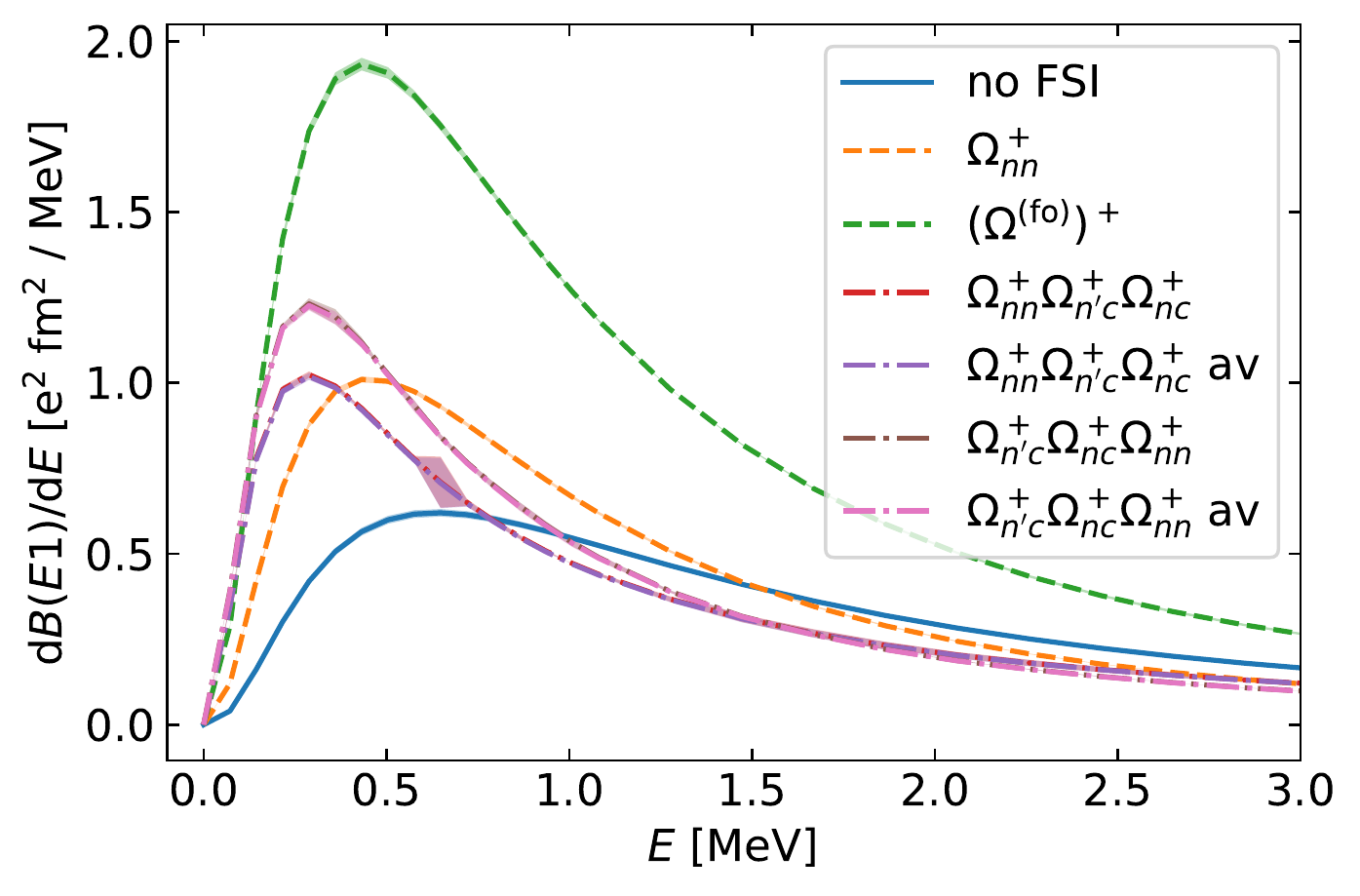}
  \includegraphics[width=0.45\textwidth]{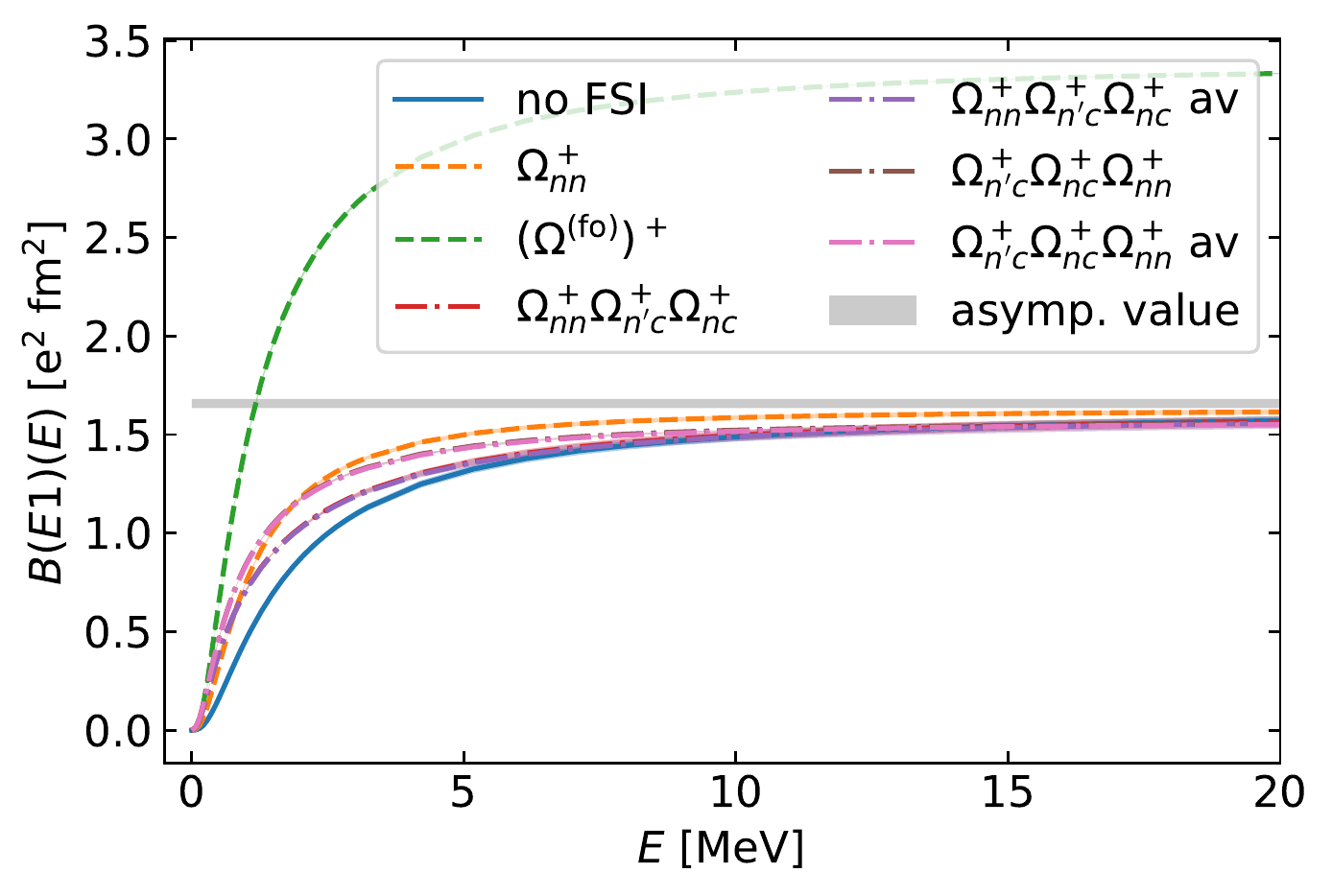}
  \caption{The left panels shows \(E1\) strength distributions of \lieleven~with different FSIs including higher-order schemes.
  The right panel shows the corresponding cumulative \(E1\) strength distributions.
  The small horizontal band again shows the expected asymptotic value for the cumulative \(E1\) strength distribution, 
  based on \(\expval{r_c^2}\) extracted from \(\mathcal{F}_c{\K{k^2}}\).
  Note that the results for \(\Omega_{nn}^\dagger \Omega_{n'c}^\dagger \Omega_{nc}^\dagger\) and for \(\frac{1}{2} \Omega_{nn}^\dagger \K{ \Omega_{n'c}^\dagger \Omega_{nc}^\dagger + \Omega_{nc}^\dagger \Omega_{n'c}^\dagger}\) are on top of each other.
  The same is true for \(\Omega_{n'c}^\dagger \Omega_{nc}^\dagger \Omega_{nn}^\dagger\) and \(\frac{1}{2} \K{ \Omega_{n'c}^\dagger \Omega_{nc}^\dagger + \Omega_{nc}^\dagger \Omega_{n'c}^\dagger} \Omega_{nn}^\dagger\).}
  \label{fig:e1_distribs_mo_combs_fsi}
\end{figure}

The \(E1\) distributions based on \(\mofod\), \(\Omega_3^\dagger\), \(\bar{\Omega}_3^\dagger\), \(\K{\Omega_3^\prime}^\dagger\), and \(\K{\bar{\Omega}_3^\prime}^\dagger\) are shown in \cref{fig:e1_distribs_mo_combs_fsi}.   Numerical uncertainties are indicated by bands, which are very narrow here.
  They were obtained by comparing the calculations with ones having roughly two thirds as many mesh points
  and a cutoff of three fourths of the original one.
The right panel of \cref{fig:e1_distribs_mo_combs_fsi} contains the cumulative distributions.
 Some of the distributions already shown in \cref{fig:e1_distribs_sg_fsi} are also included for comparison.

A striking feature of Fig.~\ref{fig:e1_distribs_mo_combs_fsi} is how much the distribution using \(\mofod\) (green dashed curve) differs from all the others: it has much more strength than any of them. It violates the non-energy-weighted sum rule by a significant margin, attaining an asymptotic value that is roughly twice as large as it should be. The deviation is not totally surprising, as this FSI operator is only an approximation to a unitary M{\o}ller operator.

In contrast, there is no large difference between the distribution using \(\Omega_3^\dagger\) (crimson dot-dashed curve) and the one just having \(nn\) FSI (orange dashed curve). 
This combination of M{\o}ller operators includes the same first-order terms in the multiple-scattering series as \(\mofod\) but is explicitly unitary. It does preserve the sum rule. Including \(n'c\) and \(nc\) interactions via a product of M{\o}ller operators moves the peak of the $E1$ strength distribution to slightly lower energy and increases the peak height slightly. This observation might show that taking products of increasing numbers of M{\o}ller operators forms a convergent approximation to the multiple-scattering
series at low energies. 

\(\bar{\Omega}_3^\dagger\) does not need to be unitary, but it gives a result that is indistinguishable from that of  \(\Omega_3^\dagger\).
This implies both that \(\bar{\Omega}_3^\dagger\)  is approximately unitary (and indeed, it fufils the sum rule well) and that  the violation of antisymmetry in \(\Omega_3^\dagger\) is small.

We also consider the operator \(\K{\Omega_3^\prime}^\dagger\), which differs from \(\Omega_{3}^\dagger\) only in the position of \(\Omega_{nn}^\dagger\). 
In  \(\Omega_3^\dagger\) it is the first factor
in the product of operators, in  \(\K{\Omega_3^\prime}^\dagger\) it is the last one. We also note that results for \(\K{\Omega_3^\prime}^\dagger\) and 
 \(\K{\bar{\Omega}_3^\prime}^\dagger\) agree excellently. Since one is unitary and the other respects \(nn\) antisymmetry this again suggests that violations of these symmetries are small in either approximation scheme. However, \(\K{\Omega_3^\prime}^\dagger\)  and \(\Omega_3^\dagger\) give somewhat different results. That difference can be taken to be an estimate of 
the remaining uncertainty in the FSI. This suggests that our 
approximation to the multiple-scattering series is not fully converged, although the uncertainty due to the approximations used for computing the FSI here is certainly smaller than the uncertainty due to NLO effects. 

Any of these combinations of three M{\o}ller operators can thus be used for a comparison with experimental data, since they are either exactly, or to a high degree, \(nn\) anti-symmetric and norm preserving.
Note that this scheme can not be easily extended to order 4 in the three-body system, since then at least one M{\o}ller operator,
which we will call \(\Omega_{ij}^\dagger\), would need to appear
two times in the product. 
Even if there are other M{\o}ller operators between the two occurrences, this would also generate an factor of \( t_{ij} G_0^{(ij)} t_{ij} G_0^{(ij)} \)
in some term\footnote{
  The full factor written more formally reads
  \begin{align*} 
    & \drint{\p}{\q} \K{ \iketbra{p,q}{k}{} \otimes \orbid \otimes \spid } t_{ij}{\K{E_p}} G_0^{(ij)}{\K{E_p}} \\
    & \quad \times \drint{\qp}{\pp} \K{ \iketbra{p',q'}{k}{} \otimes \orbid \otimes \spid } t_{ij}{\K{E_{p'}}} G_0^{(ij)}{\K{E_{p'}}} \,.
  \end{align*}
} due to the identity terms in the M{\o}ller operators between the two occurrences.
This would be unphysical, as \(t_{ij}\) fully iterates the \(ij\) interaction and therefore the same t-matrix should not be
applied two times directly subsequently with only Green's functions in between.
That such a ``doubling'' is not allowed can also be seen from the expression for the multiple-scattering series in \cref{eq:mss}.

\subsection{Comparison with other theoretical results and with experimental data}\label{ssec:cmp_expm}

\begin{figure}[htb]
  \centering
  \includegraphics[width=0.9\textwidth]{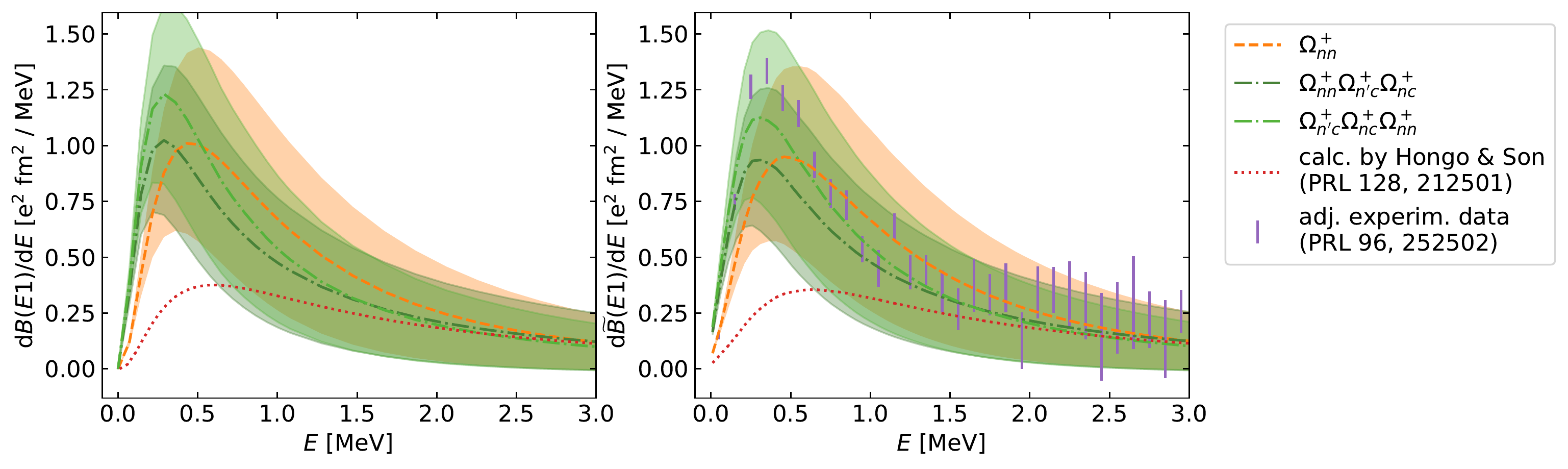}
  \caption{
  The left panels shows our results for the \(E1\) strength distribution in comparison with the universal curve by
  Hongo and Son~\cite{Hongo:2022sdr}.
  The right panel shows our results and the result of Hongo and Son folded with the detector resolution in comparsion with the
  experimental data from Nakamura et al. \cite{Nakamura:2006zz} (adjusted to the current \(S_{2n}\) value).
  The uncertainty bands show the estimated uncertainties of the leading-order EFT results.
  The uncertainty stemming from approximations of the multiple-scattering series by products of M{\o}ller operators
  can be estimated by comparing the curve using \(\Omega_{nn}^\dagger \Omega_{n'c}^\dagger \Omega_{nc}^\dagger\)
  with the one using \(\Omega_{n'c}^\dagger \Omega_{nc}^\dagger \Omega_{nn}^\dagger\).}
  \label{fig:e1_distribs_expm_uc_cmp}
\end{figure}

We proceed by comparing our results with experimental data from Ref. \cite{Nakamura:2006zz}. 
Figure~\ref{fig:e1_distribs_expm_uc_cmp} shows our results with different implementations of FSI through combinations of M{\o}ller operators.
In the left panel the results of our calculations are plotted, in the right panel these theoretical distributions folded with the detector response are shown in comparison with the experimental data\footnote{
  Within the folding the finite energy resolution as well as the finite angular resolution reported in Ref. \cite{Nakamura:2006zz}
  are taken into account.
  Moreover note that the extraction of the \(E1\) strength from the differential cross section depends on the virtual photon number
  and thereby also a dependency on the two-neutron separation energy \(S_{2n}\) enters.
  Ref. \cite{Nakamura:2006zz} from 2006 used $S_{2n}=300$ keV.
  The current value is approximately 369 keV \cite{Wang:2021xhn}.
  Therefore, we reextracted the \(E1\) curve using the current value.
  This reextraction is mainly relevant in \(E < 1\) MeV region, where it changes the peak height by approximately 10 \%.
}.
The three results differing in FSIs all have in common that \(nn\) FSI, which turned out to be rather important, is taken into account: the orange dashed line is the result 
for \(nn\) FSI alone, while the light green and dark green dot-dashed curves are two different orderings of the three possible M{\o}ller operators for this system. The difference between
the light and dark green curves can thus be taken as an estimate of the uncertainty in our approach. 
Bands indicating the uncertainties due to truncating the EFT at leading order are also shown.
We estimated those uncertainties using
\begin{equation}
  \Delta{\K{ \frac{\dd{B{\K{E1}}}}{\dd{E}} }} = \frac{\dd{B{\K{E1}}}}{\dd{E}} \sqrt{ \frac{E}{E^*{\K{{}^9\mathrm{Li}}}} } \,,
\end{equation}
whereby \(E^*{\K{{}^9\mathrm{Li}}}=2.7\) MeV is the excitation energy of the \(^9\)Li core, which is the lowest scale of omitted physics. The figure clearly shows that, in this leading-order calculation, the EFT uncertainties are larger than the uncertainties due to the treatment of FSI.

The prediction of Hongo and Son~\cite{Hongo:2022sdr} agrees well with experimental data and our results at higher energies, but has
far too little strength at low energies.
The doubtful applicability of it to this halo nucleus thereby manifests itself in a low-energy discrepancy from experimental data.

Our different distributions, which all take \(nn\) FSI into account, show qualitative agreement with the experimental values.
In the case of the height and width of the low-energy peak there are some discrepancies, which depend also on the concrete FSI approximation scheme.
That using solely \(nn\) FSI can lead to good agreement\footnote{
  However, note that the model of \lieleven~ from Ref. \cite{Esbensen:1992qbt} empolyed along with others in Ref. \cite{Esbensen:2007ir} yields an \(S_{2n}\) of 200 keV in contrast to the current experimental value of 369 keV.
} with experimental data can be also seen in Refs. \cite{Esbensen:2007ir,Nakamura:2006zz}.

We conclude that our leading-order calculation of the \(E1\) strength distribution of \lieleven~agrees reasonably well
with experimental data.
The FSI approximation technique based on products of M{\o}ller operators has proven to be useful,
in particular because it provides insight into the role of different FSIs.


\section{Conclusion}
\label{sec:conclusion}

In this work we calculated the \(E1\) strength distribution of the two-neutron halo nucleus \(^{11}\)Li using 
a three-body description in Halo effective field theory (EFT) at leading order.
We investigated the role of the final-state interactions (FSI) and found that they influence the shape of the distribution
significantly.
The results show that \(nn\) FSI is the most important single FSI.
We also investigated approximations to the full multiple-scattering series that determines the FSI.
Including all possible FSIs via a first-order treatment of their t-matrices
leads to large unitarity violations, which become manifest in large violations of the non-energy-weighted sum rule.
Therefore, we propose a unitary approximation scheme based on products of M{\o}ller scattering operators.
We were able to verify the expected compliance with the sum rule.
In computations up to third order in the t-matrix, the dominance of \(nn\) FSI was confirmed.

We have provided expressions for the \(E1\) distribution with FSI included that are suitable for application to other Borromean \(2n\) halos.
In future studies these could be computed in this framework and compared with experimental data.  
Moreover, the convergence pattern of the FSI approximations would be an interesting aspect for further studies.
These patterns should also be compared to exact calculations of the
three-body scattering state like in Ref. \cite{Kikuchi:2013ula} and alternative approaches including
full FSI effects such as the Lorentz integral transform
method~\cite{Efros:2007nq}.

The comparison of the results with experimental data showed good agreement, given that we carried out the calculation only to leading order in Halo EFT. At next-to-leading order (NLO) the impact of low-energy \(p\)-wave resonances in \({}^{10}\)Li will appear as perturbative corrections to both the initial-state bound \({}^{11}\)Li wave function and the FSI. The \(nn\) and \(nc\) effective ranges are also both an NLO effect. 

This control over final-state interactions allows us to investigate the impact of different assumptions about the \(^9\)Li-\(nn\) dynamics on the E1 strength distribution. 
We showed that a description taking all spins into account and using both \(s\)-wave \(nc\) interaction channels (\(s_c-1/2\) and \(s_c+1/2\))
at the same strength yields a good leading-order description of the \(E1\) strength.
Conveniently, such a calculation is equivalent to a calculation with only neutron spins included and therefore can be recast as a calculation with a spinless core. 
We also provided the formalism for a calculation in which only the (\(s_c+1/2\)) channel has the low-energy enhancement that leads to the \({}^{10}\)Li virtual
state. Such a calculation significantly underpredicts the data.
Finally, we compared to the EFT calculation of Hongo and
Son~\cite{Hongo:2022sdr}, which is based on different assumptions
about the underlying scales of the $nnc$ system. In particular,
it assumes that the neutron-\({}^9\)Li interaction is subleading. 
The corresponding prediction disagrees with data in the range from $E=0$--$1$ MeV by a factor of three to four. Since all calculations are adjusted to have the same \(S_{2n}\) of \({}^{11}\)Li, the differences in the \(E1\) distribution genuinely reflect the different assumptions about the \(nc\) subsystem dynamics.

\acknowledgments

We thank T. Nakamura for providing the \(E1\) strength data of Ref. \cite{Nakamura:2006zz}
and T. Aumann for useful discussions.
M.G. and H.W.H. acknowledge support by Deutsche Forschungsgemeinschaft
(DFG, German Research Foundation) - Project-ID 279384907 - SFB 1245.
H.W.H. has been supported by the German Federal Ministry of
Education and Research (BMBF) (Grant No. 05P21RDFNB).
B.A. acknowledges support by the Neutrino Theory Network Program (Grant No. DE-AC02-07CH11359).
D.R.P. was supported by the US Department of Energy,
contract DE-FG02-93ER40756.

\newpage
\appendix


\section{\texorpdfstring{Mapping the calculation with two \(\boldsymbol{nc}\) interaction channels onto the spinless calculation}
  {Mapping the calculation with two nc interaction channels onto the spinless calculation}}
\label{ap:equiv_one_channel_two_channels}

In this section we consider \(2n\) halos where the core as well as the whole halo have spin \(s_c\).
We show that a leading-order description of these systems using two \(nc\) interaction channels (\(s_c - 1/2\) and \(s_c+1/2\))
is equivalent to a description with only the neutron spins taken into account and thereby having necessarily only one \(nc\) interaction channel.

\textit{Definitions:}
The spin states describing such a system seen from the core as spectator or a neutron as spectator are given by
\begin{align}
  \iket{ \xi_c^{\K{\sigma;s_c,M}} }{c} &= \iket{\K{\oh,\oh}\sigma,s_c;s_c,M}{c} \quad \sigma \in \{0,1\} \,, \\
  \iket{ \xi_n^{\K{\tau;s_c,M}}   }{n} &= \iket{\K{\oh,s_c} s_c+\tau \oh,\oh;s_c,M}{n} \quad \tau \in \{-1,+1\} \,.
\end{align}
The corresponding projection operators are
\begin{align}
  P_c^{(\sigma)} &= \sum_M \iketbra{\xicmi{\sigma}}{c}{} \,, \\
  P_n^{(\tau)}   &= \sum_M \iketbra{\xinmi{\tau}}{n}{} \,.
\end{align}

\textit{Statement:}
Having these definitions at hand we can now state that the mapping can be made if the Hamilton operator has the structure
\begin{align}
  H_0    &= H_0^{(\mathrm{spatial})} \otimes \spid = H_0^{(\mathrm{spatial})} \otimes \K{ \pcz + \pco } \,, \\
  V_{nn} &= V_{nn}^{(\mathrm{spatial})} \otimes \pcz \,, \\
  V_{nc} + V_{n'c} &= \K{ V_{nc}^{(\mathrm{spatial})} + V_{n'c}^{(\mathrm{spatial})} } \otimes \K{ \pnm + \pnp } \label{eq:V_nc_sum}
\end{align}
and
\begin{equation}\label{eq:proj_op_rel}
  \pcz + \pco = \pnm + \pnp
\end{equation}
holds.
\Cref{eq:V_nc_sum} means that the spatial/momentum-space part of the \(nc\) interaction has to be the same in \(s_c-1/2\) and \(s_c+1/2\).

More specifically, in this case the Schrödinger equation can be decoupled into one in the \(\pcz\)-space and one in the
\(\pco\)-space\footnote{
  This is because \(V_{nc} + V_{n'c} = \K{ V_{nc}^{(\mathrm{spatial})} + V_{n'c}^{(\mathrm{spatial})} } \otimes \K{ \pcz + \pco}\) holds then.
}:
\begin{equation}
  H = H^{(0)} + H^{(1)} = H^{(\mathrm{spatial};0)} \otimes \pcz + H^{(\mathrm{spatial};1)} \otimes \pco
\end{equation}
While the \(\pco\)-space Schrödinger equation misses an \(nn\) interaction, the \(\pcz\)-space one is equivalent to a calculation with \(s_c=0\).
This equivalent equation has the Hamilton operator
\begin{align}\label{eq:H0_factorization}
  H^{(0)} &= H^{(\mathrm{spatial};0)} \otimes \pcz \,, \\
  H^{(\mathrm{spatial};0)} &= H_0^{(\mathrm{spatial})} + V_{nc}^{(\mathrm{spatial})} + V_{n'c}^{(\mathrm{spatial})} + V_{nn}^{(\mathrm{spatial})} \,. \label{eq:Hsp0}
\end{align}

\textit{Sketch of the proof:}
The relation for the projection operators given in \cref{eq:proj_op_rel} can be verified by
inserting
\begin{equation}
  \ket{\xicmi{\sigma}} = \sqrt{2\sigma + 1} \sqrt{2s_c}   \begin{Bmatrix} 1/2 & s_c & s_c-1/2 \\ s_c & 1/2 & \sigma \end{Bmatrix} \ket{\xinmi{-}}
    + \sqrt{2\sigma + 1} \sqrt{2s_c+2} \begin{Bmatrix} 1/2 & s_c & s_c+1/2 \\ s_c & 1/2 & \sigma \end{Bmatrix} \ket{\xinmi{+}}
\end{equation}
into \(\pcz + \pco\) and using the ortho-normality relations for the Wigner-6j symbols, as they can be found in Ref. \cite{Varshalovich1989}.
The other pillar of the proof is to show that in a leading-order calculation with \(s_c = 0\) the Hamilton operator
has indeed the form
\begin{align}\label{eq:H_sep}
  H &= H^{(\mathrm{spatial})} \otimes P \,, \\
  H^{(\mathrm{spatial})} &= H_0^{(\mathrm{spatial})} + V_{nc}^{(\mathrm{spatial})} + V_{n'c}^{(\mathrm{spatial})} + V_{nn}^{(\mathrm{spatial})} \,.
\end{align}
This can be shown by introducing the spin states \(\iket{\xi_c}{c}\) for the \(nn\) interaction channel and \(\iket{\xi_n}{n}\) for the
\(nc\) interaction channel. As the core has here spin zero and the overall spin is zero, the only allowed \(nn\) spin configuration is zero.
Thereby the two states are equal up to a sign and the Hamilton operator for \(s_c = 0\) takes indeed this form.

\section{\texorpdfstring{Explicit relations for \(\boldsymbol{\mathcal{A}^{(1)}}\)}{Explicit relations for A1}}
\label{ap:mc_A_1}
\newcommand{\ft}{\widetilde{f}}

In the following we give equations suitable for evaluating \(\mathcal{A}^{(1)}\), which is defined in \cref{eq:def_maf1}:
\begin{align}
    \maf{1}{\lb,\lbb;\mu;\bar{s},M}{p,q} &= 
    - \sqrt{2\lb+1} \sqrt{2\lbb+1} \begin{pmatrix} \lbb & 1 & \lb \\ 0 & 0 & 0 \end{pmatrix} 
    \sqrt{\pi} 
    \K{ p \bar{f}_{\lbb}{\K{p,q}} - \frac{1}{2} q \bar{f}_{\lb}{\K{p,q}} } 
    \nonumber \\ 
    &\quad \times \imel{c}{\K{\bar{s},\frac{3}{2}}\frac{3}{2},M}{P_{\xi_n}}{\xi_c^{(M)}}{c} \,, \label{eq:fr_maf1}
  \end{align}
  whereby the round brackets with six arguments denote a Wigner-3j symbol.
  Equations for \(\bar{f}\) and its ingredients are given below:
  \begin{align}
    \bar{f}{\K{ p,q,x\coloneqq \cos{\K{\theta_{\vpq}}} }} &\coloneqq 
    \frac{1}{ \kcnq{\K{p,q,x}} } \rint{\ptp} g_0{\K{\kcnp{\K{p,q,x}}}} \tau_{nc}{\K{\kcnp{\K{p,q,x}}}} g_0{\K{\ptp}} G_0^{(nc)}{\K{\ptp;E_{\kcnp{\K{p,q,x}}}}} \nonumber \\
    &\quad \times \sqrt{\pi} \K{ -\ptp \ft_1{\K{\ptp,\qtp}} - \frac{A}{A+1} \qtp \ft_0{\K{\ptp,\qtp}} } \Big|_{\qtp = \kcnq{\K{p,q,x}}} \,,\\  
    \ft{\K{ \ptp,\qtp,\tilde{x}' \coloneqq cos{\K{\theta_{\ptpv,\qtpv}}} }} &\coloneqq \frac{ f{\K{\kncp{\K{\ptp,\qtp,\tilde{x}'}},\kncq{\K{\ptp,\qtp,\tilde{x}'}}}} }{ \kncq{\K{\ptp,\qtp,\tilde{x}'}} } \,, \label{eq:def_ft} \\
    f{\K{p, q}} &= \ci \sqrt{\frac{1}{4\pi}} e Z_c \frac{2}{A+2} \K{ \partial_{\qt} \Psi_c{\K{p,\qt}} }\big|_{\qt=q} \,,
  \end{align}
  whereby the functions \(\v{\kappa}_{ijk}\) (\(i,j \in \{n,c\}\) and \(k \in \{p,q\}\)) are defined in Ref. \cite{Gobel:2019jba}.
  Furthermore, we used the following generic definition of a function \(f_i{\K{p,q}}\) via
  \begin{equation}\label{eq:gen_pw_projection}
    f_i{\K{p, q}} \coloneqq \int \dd{x} P_i{\K{x}} f{\K{p, q, x}} \,.
  \end{equation}
  The \(i\)-th Legendre polynomial is denoted by \(P_i\).

  In order to obtain these expressions inter alia the following relations and techniques were employed:
  \begin{itemize}
    \item relation for expressing \(\y{l}{m}{\v{a}+\v{b}}\) using \(\y{l}{m}{\v{a}}\) and \(\y{l}{m}{\v{b}}\) (see, e.g., Ref. \cite{gloeckle2012quantum}),
    \item relations for recoupling the Jacobi momenta (see, e.g., Ref. \cite{Gobel:2019jba}),
    \item expansion of functions in terms of Legendre polynomials and expressing Legendre polynomials in terms of \(\cywa{l,l}{0,0}\) (see, e.g., Ref. \cite{gloeckle2012quantum}),
    \item relation for the integral of three spherical harmonics (see, e.g., Ref. \cite{Varshalovich1989}).
  \end{itemize}

  \section{\texorpdfstring{Explicit relations for \(\boldsymbol{\mathcal{A}^{(2)}}\)}{Explicit relations for A2}}
  \label{ap:mc_A_2}
    
  \newcommand{\ftt}{\widetilde{f}^{(2)}}
  \newcommand{\fbbt}{\overline{\overline{f}}^{(2)}}
  \newcommand{\fbbtb}{\overline{\overline{f}}^{(2b)}}
  \newcommand{\fbt}{\overline{f}^{(2)}}
  \newcommand{\fbtb}{\overline{f}^{(2b)}}
  \newcommand{\ttau}{\widetilde{\tau}}

  We give an expression for \(\mathcal{A}^{(2)}\), which is defined in \cref{eq:def_maf2}:
  \begin{align}
    \maf{2}{\lb,\lbb;\mu;\bar{s},M}{p,q} &= \K{-1}^{\bar{l}} 
    \sqrt{\pi} \sqrt{2\lbb+1} \sqrt{2\lb+1} 
      \begin{pmatrix} \lbb & 1 & \lb \\ 0 & 0 & 0 \end{pmatrix} \K{ p \fbt_{\lbb}{\K{p,q}} - \frac{1}{2} q \fbt_{\lb}{\K{p,q}} } \nonumber \\
    &\quad \times \imel{c}{\K{\bar{s},\frac{3}{2}} \frac{3}{2}, M}{ \pmospin P_{\xi_n} \pmospin }{\xi_c}{c} \,,
  \end{align}
  Also here \cref{eq:gen_pw_projection} applies. The function \(\fbt\) is given by
  \begin{align}\label{eq:def_fbt}
    \fbt{\K{p,q,x \coloneqq \cos{\theta_{\vpq}} }} &\coloneqq \frac{2\pi}{\kcnq{\K{p,q,x}}} \rint{\ptp} \ttau_{nc}{\K{\kcnpb{p,q,x}}} g_0{\K{\ptp}} G_0^{(nc)}{\K{\ptp;E_{\kcnpb{p,q,x}}}} \nonumber \\
    &\quad \times \K{ \ptp \fbbt_1{\K{\ptp, \kcnqb{p,q,x}}} 
      - \frac{\kcnqb{p,q,x}}{A+1} \fbbt_0{\K{\ptp, \kcnqb{p,q,x}}} } \,,
  \end{align}
  whereby the short-hand notation
  \begin{equation}
    \ttau_{nc}{\K{p}} \coloneqq g_0{\K{p}} \tau_{nc}{\K{p}}
  \end{equation}
  is used.
  \Cref{eq:def_fbt} can be rewritten into
  \begin{align}
    \fbt{\K{p,q,x}} &= \frac{g_0{\K{\bar{p}}}}{2\pi^2}  \frac{1}{a_{nc}^{-1}+\ci\bar{p}} \nonumber \\
    &\quad \times \bigg[ \int_0^\Lambda \dd{p'} \frac{ p'^2 \fbtb{\K{p';p,q,x}} - \bar{p}^2 \fbtb{\K{\bar{p};p,q,x}} }{\bar{p}^2 - p'^2} 
    - I{\K{\bar{p};\Lambda}} \bar{p} \fbtb{\K{\bar{p};p,q,x}} \bigg]
    \bigg|_{\bar{p} = \kcnpb{p,q,x}} \,.
  \end{align}
  Hereby the relations
  \begin{align}
    \fbtb{\K{\ptp; p,q,x}} &\coloneqq \frac{2\pi}{\kcnq{\K{p,q,x}}} \K{ \ptp \fbbt_1{\K{\ptp, \kcnqb{p,q,x}}} 
    - \frac{\kcnqb{p,q,x}}{A+1} \fbbt_0{\K{\ptp, \kcnqb{p,q,x}}} } \,, \\
    I{\K{\bar{p};\Lambda}} &\coloneqq \K{\frac{\ci \pi}{2} - \frac{1}{2} \ln{\K{\frac{\Lambda + \bar{p}}{\Lambda -\bar{p}}}}} g_0{\K{\bar{p}}} \,.
  \end{align}
  are employed.
  Furthermore the definition
  \begin{align}
    \fbbt{\K{p,q,x\coloneqq \cos{\theta_{\vpq}}}} &\coloneqq \ttau_{nc}{\K{\knnppb{p,q,x}}} \rint{\ptpp} g_0{\K{\ptpp}} 
      G_0^{(nc)}{\K{\ptpp, E_{\knnppb{p,q,x}}}} \nonumber \\
    &\quad \times  \frac{\sqrt{\pi}}{ \knnqpb{p,q,x}} \K{-\ptpp \ft_1{\K{\ptpp, \knnqpb{p,q,x}}} - \frac{A}{A+1} \knnqpb{p,q,x} \ft_0{\K{\ptpp,\knnqpb{p,q,x}}}} \,
  \end{align}
  holds.
  This function can be rewritten in a similar way:
  \begin{align}
    \fbbt{\K{p,q,x}} &= \frac{g_0{\K{\bar{p}}}}{2\pi^2} \frac{1}{a_{nc}^{-1}+\ci\bar{p}} \nonumber \\
    &\quad \times \bigg[ \int_0^\Lambda \dd{p'} \frac{ p'^2 \fbbtb{\K{p';p,q,x}} - \bar{p}^2 \fbbtb{\K{\bar{p};p,q,x}} }{\bar{p}^2 - p'^2} 
    - I{\K{\bar{p};\Lambda}} \bar{p} \fbbtb{\K{\bar{p};p,q,x}} \bigg]
    \bigg|_{\bar{p} = \knnppb{p,q,x}} \,,\\
    \fbbtb{\K{\ptpp;p,q,x}} &\coloneqq \frac{\sqrt{\pi}}{ \knnqpb{p,q,x}} \K{-\ptpp \ft_1{\K{\ptpp, \knnqpb{p,q,x}}} - \frac{A}{A+1} \knnqpb{p,q,x} \ft_0{\K{\ptpp,\knnqpb{p,q,x}}} } \,.
  \end{align}
  Hereby \(\ft\) is the one from \cref{eq:def_ft}.

  Also for obtaining these expressions the relations and techniques listed in \cref{ap:mc_A_1} were employed.

\section{Sums over the quantum numbers of the final state and their convergence}\label{ap:qnsum_trunc}

In this appendix we briefly discuss the handling of the partial waves of the final states.
\cref{eq:one_mo_mel,eq:two_mo_mel,eq:three_mo_mel} show that only those terms
directly proportional to \(\maf{1}{\lb,\lbb;\mu;\lsb,M}{p,q}\) or \(\maf{2}{\lb,\lbb;\mu;\lsb,M}{p,q}\)
are non-zero for multiple combinations of final-state quantum numbers \(\lb\), \(\lbb\).
In contrast to that, other terms are only non-zero for \(\lb = 0\) together with \(\lbb = 1\).
The expressions for \(\maf{1}{\lb,\lbb;\mu;\lsb,M}{p,q}\) and \(\maf{2}{\lb,\lbb;\mu;\lsb,M}{p,q}\) in \cref{ap:mc_A_1,ap:mc_A_2} show that these are already non-vanishing
if \(\lbb - 1 \leq \lb \leq \lbb +1\).
Using this condition restricts the sum over \(\lb\) for a given \(\lbb\) to a finite number of 
terms, while the sum over \(\lbb\) stays in principle unrestricted.
Therefore we truncate the sum over \(\lbb\) at \(\lbb_{\mathrm{max}}\) (inclusive).
We usually use \(\lbb_{\mathrm{max}} = 5\), because the relative changes between the results based on 
\(\lbb_{\mathrm{max}} = 3\) and those based on \(\lbb_{\mathrm{max}} = 5\) are smaller than 5 \% measured in terms of the former.
(In fact, in the \(E < 3\) MeV region, which we show in most plots, the relative change is below 2.5 \%.)
In the case of the quantum number \(\mu\) the sum runs from -1 to 1, and we use the fact that the matrix element
is independent of \(\mu\) in order to reduce the numerical costs.
The spin of the \(nn\) system in the final state can generally be 0 or 1, while in the case of some terms only 0 is possible.
Moreover, sometimes cancellations emerge for certain values naturally because of the nature of the equations.
E.g., in the case of \(\Omega_3^\dagger\) the partial wave \(s = 1 \, \land \, l = 0\) has in principle a non-vanishing contribution, as \(\Omega_3^\dagger\) does not
commute with \(\pmo\), while in the case of \(\bar{\Omega}_3^\dagger\) this contribution is vanishing due to the \(nn\) antisymmetry of the operator.

\begin{figure}[H]
  \centering
  \includegraphics[width=0.7\textwidth]{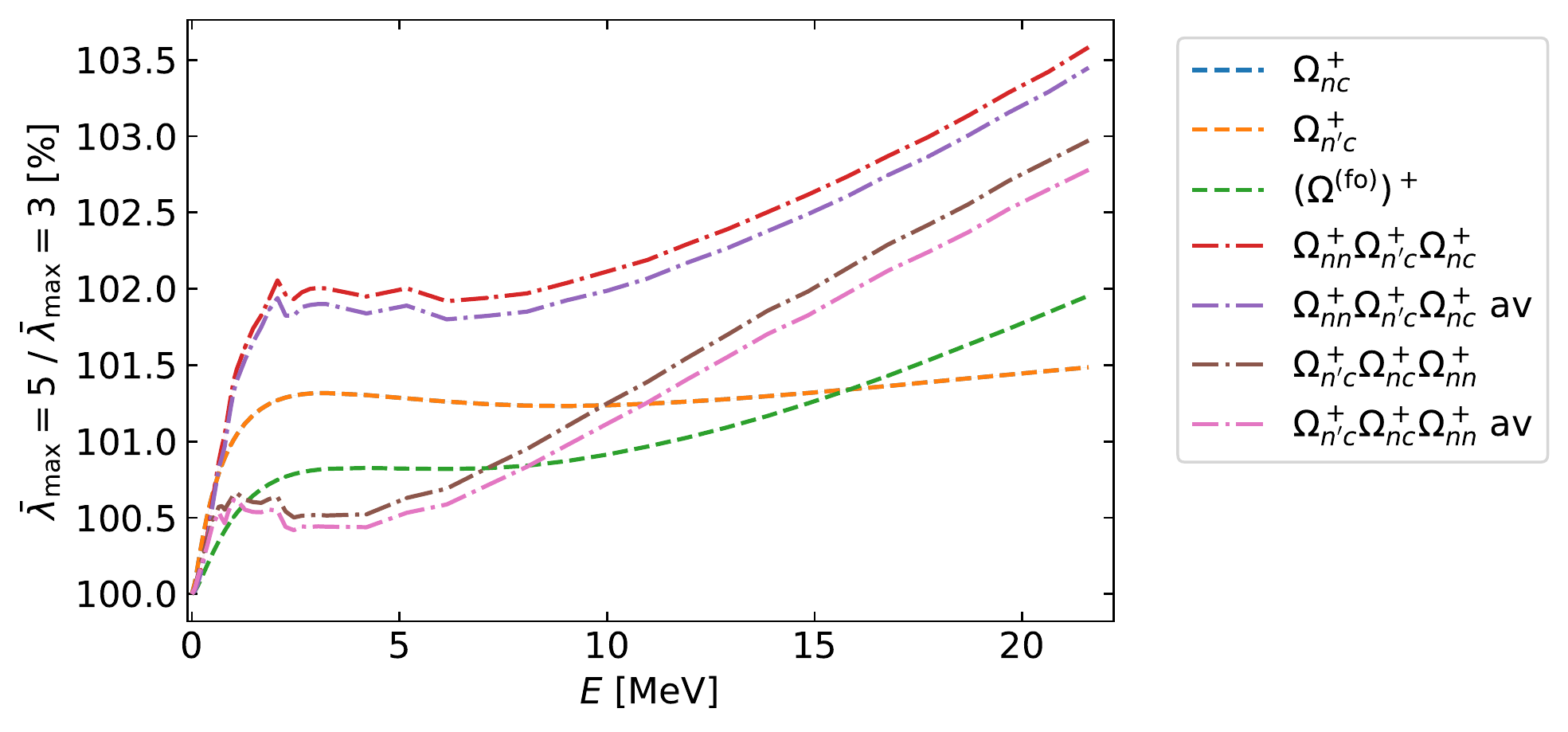}
  \caption{Quotients of \(E1\) strength distributions with \(\bar{\lambda}_\mathrm{max} = 5\) and with
  \(\bar{\lambda}_\mathrm{max} = 3\) differing in the FSI treatment.}
  \label{fig:estimate_lbm_trunc_error}
\end{figure}

Finally, we present numerical data on the error orginating from the truncation in the quantum number \(\lbb\) at \(\bar{\lambda}_{\mathrm{max}}\).
\Cref{fig:estimate_lbm_trunc_error} shows the quotients of distributions obtained with \(\bar{\lambda}_\mathrm{max} = 5\) and
\(\bar{\lambda}_\mathrm{max} = 3\).

In the case of the shown distributions the relative changes are smaller than 5 \%.
Given the significant EFT uncertainty bands at leading order this is sufficient precision.

\bibliographystyle{apsrev4-2}

\end{document}